\documentclass[aps,prb,twocolumn,superscriptaddress,longbibliography]{revtex4-1}

\pdfoutput=1
\usepackage[T2A,T1]{fontenc}
\usepackage[latin9]{inputenc}
\usepackage{graphicx}
\usepackage{graphics}
\usepackage{dsfont}
\usepackage{bm}
\usepackage{amsmath}
\usepackage{caption}

\usepackage{verbatim}
\usepackage{textcomp}
\usepackage{amssymb}
\usepackage{esint}
\usepackage{units}

\makeatletter

\newcommand{\abs}[1]{ \left\lvert#1\right\rvert} % absolute value: single vertical bars

\@ifundefined{showcaptionsetup}{}{%
 \PassOptionsToPackage{caption=false}{subfig}}
\usepackage{subfig}
\makeatother

\begin{document}

\title{Non-local quantum gain facilitates loss compensation and plasmon amplification in graphene hyperbolic
metamaterials}

\author{Illya I. Tarasenko}
\affiliation{The Blackett Laboratory, Department of Physics, Imperial College London,
London SW7 2AZ, United Kingdom}

\author{A. Freddie Page}
\affiliation{Dyson School of Design Engineering, Imperial College London,
London SW7 2DB, United Kingdom}

\author{Joachim M. Hamm}
\affiliation{The Blackett Laboratory, Department of Physics, Imperial College London,
London SW7 2AZ, United Kingdom}

\author{Ortwin Hess}
\affiliation{The Blackett Laboratory, Department of Physics, Imperial College London,
London SW7 2AZ, United Kingdom}

\begin{abstract}
Graphene-based hyperbolic metamaterials have been predicted to transport evanescent fields with extraordinarily large vacuum wave-vectors. It is particularly at much higher wave vector values that the commonly employed descriptional models involving structure homogenization
and assumptions of an approximatively local graphene conductivity start breaking down. Here, we combine a non-local quantum conductivity model of graphene with an exact mathematical treatment of the periodic structure in order to develop a tool-set for determining the hyperbolic behavior of these graphene-based hyperbolic metamaterials. The quantum conductivity model of graphene facilitates us to predict the plasmonic amplification in graphene sheets of the considered structures. This allows us to reverse the problem of Ohmic and temperature losses, making this simple yet powerful arrangement practically applicable. We analyze the electric field distribution inside of the finite structures, concluding that Bloch boundary solutions can be used to predict their behavior. With the transfer matrix method we show that at finite temperature and collision loss we can compensate for losses, restoring imaging qualities of the finite structure via an introduction of chemical imbalance.
\end{abstract}
\maketitle

\section{Introduction}

Graphene-based hyperbolic metamaterials have been predicted to transport
evanescent fields with large vacuum wave-vectors \citep{Shung1986,Yan2012,Iorsh2013,Sreekanth2013,MohamedA.K.Othman2013,Poddubny2013,Ferrari2015,Chang2016, Novoselov2016}.
This feature has immense practical importance, as it opens up possibilities
of manipulating wavevectors outside of the light cone, that have been
inaccessible before \citep{Veselago1968,Pendry2000}. This is a desirable
property in fields such as imaging \citep{Belov2006,Salandrino2006,Wood2006,Rho2010,Andryieuski2012},
control of spontaneous emission \citep{Galfsky2015,Tumkur2011,Iorsh2013}, nano-lithography
\citep{Xiong2009,Ferrari2015}, and others \citep{Ferrari2015,Smolyaninov2011}.
It has also been reported that due to the properties of the carrier
system in graphene, the response of such metamaterials can indeed be dynamically
tuned and switched between elliptic and hyperbolic regimes of
operation \citep{Iorsh2013,MohamedA.K.Othman2013,Dai2015a}.

Concurrently, moving deeper into the sub-wavelength region, the
models, currently used to predict properties of such systems, lose
their applicability: the medium homogenization approximations fail
when the wavelengths are comparable to the unit cell thickness \citep{Orlov2011,Orlov2014}; non-local effects have a significant influence on the graphene properties for the excitations with
wavevectors approaching the Dirac cone \citep{Page2015}. At the
same time, Ohmic losses, which are an inherent problem to all hyperbolic
metamaterials (HMMs) with conductive sheets \citep{Savelev2013}, are an important issue
for graphene-based HMMs (GHMMs) too. However, calculation of losses had been limited due to the approximations of the local graphene conductivity model.

In this study we apply a non-local quantum conductivity model,
derived from irreducible polarizability of a particle-hole plasma to GHMMs.
This facilitates exact in RPA calculations of the carrier distributions
and interactions at arbitrary temperatures and levels of doping \citep{Page2015,Pyatkovskiy2009a,Wunsch2006,Hwang2007}.
Due to the non-locality of the model, we are able to perform exact calculations
of the response of graphene to the excitations with in-plane wave-vectors
of energies up to the edge of the Dirac cone, which has previously been impossible at the basis of the local conductivity models. It also provides a description
of graphene in the photo-inverted state, where plasmons, coupled to the excited charge carriers, will stimulate electron-hole pair recombination with consequential amplification of
the triggering plasmons, uncovering a new property of graphene in the context of GHMMs \citep{Page2015,Pyatkovskiy2009a,Watanabe2013,Dubinov2011}.

Further, we couple this conductivity model to a periodic structure
description, treating such an arrangement as a photonic crystal, to calculate
and identify where the metamaterial exhibits elliptic or
hyperbolic properties. Due to the linearity of the Dirac Cone, the
proposed band structure is of a general nature, describing simultaneously
the behavior of the material under all physical values of excitation
frequencies and doping levels for the separate cases of passive and
active graphene. This generalized treatment eliminates the necessity
to study individual regions of doping and excitation values, as it
covers the full parameter space. We propose a scheme, in which
we can calculate the effective optical parameters of the structure
exactly, avoiding, in particular, homogenization techniques, which
have a limited validity. We also provide a set of analytic tools and
treatments, which could be used for an efficient characterization
of graphene-based HMMs, as well as any HMMs where the conductive layers
are a two-dimensional material.

We use the developed methods in order to analyze the behavior of the structure when the graphene sheets are doped, photo-inverted and at realistic temperatures. We show that in such an arrangement the stimulated plasmonic emission occurs in a frequency band of regulated width, as opposed to the line of the single graphene sheet, suspended in air. Further analysis shows that we can compensate and overcompensate Ohmic losses, opening an avenue to removing an important obstacle to a practical realization of hyperbolic metamaterials.

Finally, we calculate the quasi-static electric field distribution in finite GHMM structures, generated by a propagating narrow spatial Gaussian field envelope. In comparison with the loss-less case, we shed light on the degrading effects of collision and temperature losses on imaging of the envelope in the passive structure. Then, putting graphene sheets in the state of optical inversion, we show the length of propagation of the field through the structure being restored due to the plasmonic gain to that of the ideal system with no losses.

\section{Quantum Conductivity model}

\begin{figure*}\subfloat[\label{fig:2.1a}]{\includegraphics[width=0.5\linewidth]{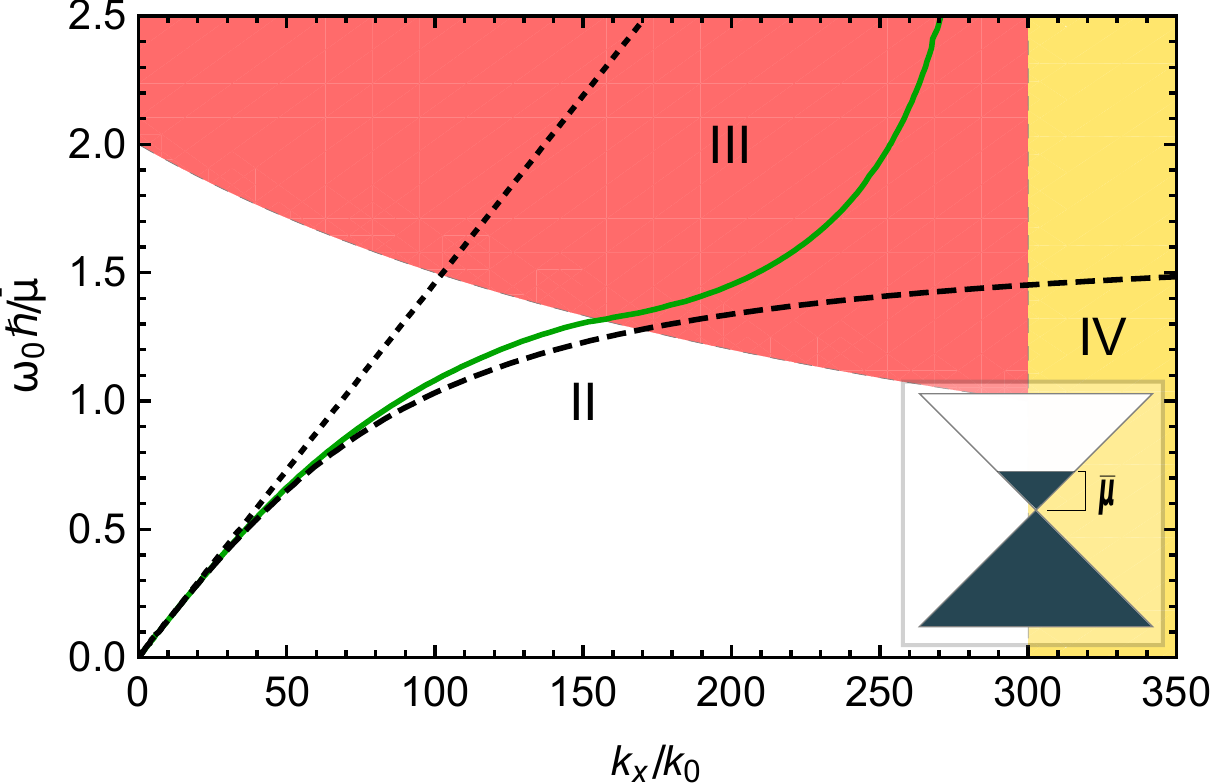}}\hfill{}\subfloat[\label{fig:2.1b}]{\includegraphics[width=0.5\linewidth]{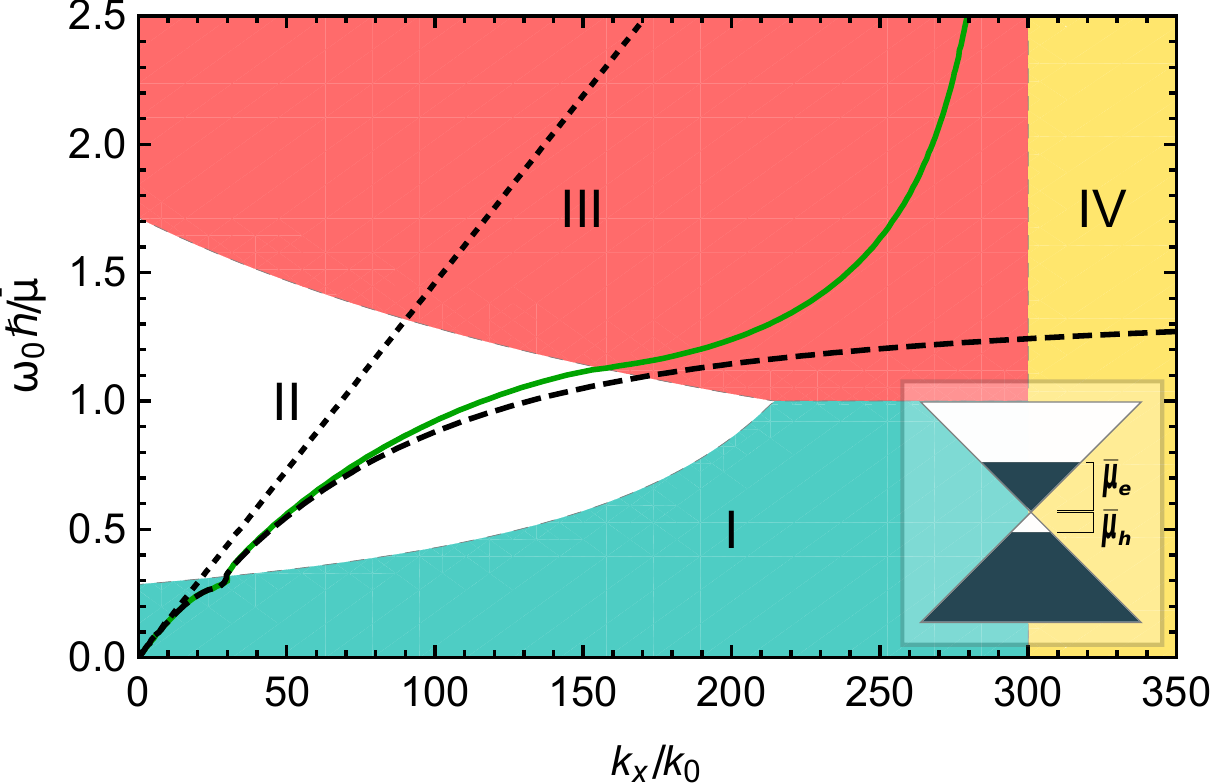}}

\caption{Plasmon frequency dispersion of a doped, air-suspended single graphene sheet without photo-inversion (a) and with photo-inversion (b). The
non-local quantum conductivity model is used for obtaining the solid
green curves, and a modified local Drude model for the dashed curves.
The excitation frequency on the vertical axis is scaled with $\omega_{pl}\hbar/\bar{\mu}$. The in-plane component of the excitation wavevector on the horizontal axis is scaled with the excitation frequency to transform the light cone and the Dirac cone into vertical lines at $k_{x}/k_{0}=1$ and $k_{x}/k_{}=c_{0}/v_{Fermi}=300$ respectively. The plasmons experience gain within region I due to inter-band electron-hole recombination; loss - within regions III
and IV due to the inter- and intra-band absorption respectively; and
propagate without loss or gain within region II due to the absence
of phase space for any inter- or intraband transitions.\label{fig:2.1}}
\end{figure*}

In order to calculate exactly the plasmon dispersion in graphene and predict plasmon amplification or attenuation, it is necessary to use a conductivity
model which is both non-local and remains valid within the regions
of inter-band processes. This requires an exact treatment of large
wavevectors and complex frequencies. This is not achieved with the local Drude \citep{Sreekanth2013} and Kubo \citep{Iorsh2013,MohamedA.K.Othman2013}
formulas%
\begin{comment}
need to check once more that the equations in references correspond
to Kubo and Drude models.
\end{comment}
, most often used in studies of graphene-based HMMs. They become increasingly
inaccurate in the parameter space of the inter- and intra-band transitions
\citep{Page2015}, limiting their applicability to scenarios of small loss and absence of gain. Therefore, in order to grasp loss compensation and amplification in graphene-based HMM, we employ a non-local quantum conductivity model, derived from the polarization function \citep{Page2015}:

\begin{equation}
\sigma_{s}(q,\omega)=\frac{i\omega e^{2}\Pi(q,\omega)}{q^{2}}\label{eq:2.1-1}
\end{equation}
where $\Pi(q,\omega)$ is the propagator of the electron-hole pairs.
For a general 2D electron gas in random phase approximation (RPA)
it is given by the Lindhard formula \citep{Stern1967,Shung1986} - see Appendix
Eq. \eqref{eq:A.1}. A closed-form analytic expression for gap-less
graphene at zero temperature has been derived \citep{Pyatkovskiy2009a},
and is presented in the Appendix A as Eq. \eqref{eq:A.2}.

The exact Eq. \eqref{eq:A.2} needs modification for the case of more
complicated distribution of electrons $n\left(\varepsilon\right)$ and
holes $\bar{n}\left(\varepsilon\right)$ (e.g. non-zero temperature,
photo-excitation). Using the fact that the Lindhard formula is linear
in the carrier distribution $n\left(\varepsilon\right)$,
the polarization function of the arbitrary non-equilibrium graphene
sheet can then be expressed in the form \citep{Page2015,Page2016}:

\begin{align}
\Pi\left(n\right) & =\Pi|_{\mu=0}^{T=0}+\underbrace{\intop_{0}^{\infty}d\varepsilon\left[\frac{\partial\Pi\left(q,\omega\right)|_{\mu=\varepsilon}^{T=0}}{\partial\varepsilon}n\left(\varepsilon\right)\right]}_{\Pi^{\left(e\right)}[n]}\nonumber \\
 & +\underbrace{\intop_{0}^{\infty}d\varepsilon\left[\frac{\partial\Pi\left(q,\omega\right)|_{\mu=\varepsilon}^{T=0}}{\partial\varepsilon}\bar{n}\left(\varepsilon\right)\right]}_{\Pi^{\left(h\right)}[\bar{n}]}\label{eq:2.2-1}
\end{align}

where $\Pi^{\left(e\right)}[n]$ and $\Pi^{\left(h\right)}[\bar{n}]$
are the individual contributions of the electrons and holes. Here,
$n\left(\varepsilon\right)$ and $\bar{n}\left(\varepsilon\right)=1-n\left(-\varepsilon\right)$
are arbitrary functions of the distribution of electrons and holes,
allowing to incorporate chemical doping, photo-excitation and temperature \citep{Page2018,Hamm}.

The plasmon dispersion of graphene is calculated using the exact in RPA model for zero temperature doped graphene sheet without and with photo-excitation, and is presented as green solid lines in the Figs. \ref{fig:2.1a} and \ref{fig:2.1b} respectively. The excitation frequency on the vertical axis is scaled with the value of chemical imbalance $\bar{\mu}$, and the in-plane wavevector on the horizontal axis is scaled with the excitation frequency. With such scaling the light cone transforms into the vertical line at $k_{x}/k_{0}=1$ and the Dirac cone transforms into the vertical line at $k_{x}/k_{0}=c_{0}/v_{Fermi}=300$. The plots without excitation frequency scaling are shown in the Figs. \ref{fig:A1a} and \ref{fig:A1b} in Appendix A.

Chemical doping of the graphene sheet causes a shift of the Fermi level away from
the Dirac point. This creates a region in phase space where inter-band processes are not possible due to insufficient excitation energies. This corresponds to the region we denote by II in the Figs. \ref{fig:2.1a} and \ref{fig:2.1b}. The excitations with energies above the Fermi level could decay through Landau damping into the electron-hole pairs with energies above the Fermi level. This corresponds to the region III in the Figs. \ref{fig:2.1a} and \ref{fig:2.1b}. Region IV in both cases corresponds to inter-band excitations (but is not of interest here, as it does not provide a mechanism for the plasmon amplification).

Optical pumping of doped graphene creates a two-component plasma of
electrons and holes. This opens up a phase space within region II,
where the plasmons couple to the inverted carrier plasma, trigger
stimulated recombination and get amplified. We denote this area as the
region I in the Fig. \ref{fig:2.1b}.

The relative sizes of regions I, II and III in the active case depend
on the value of doping and imbalance between the chemical potentials
of electrons and holes (illustrated in the inserts of the Figs.
\ref{fig:2.1a} and \ref{fig:2.1b}). For example, the Fermi level is
at zero for intrinsic photo-excited graphene ($\mu_{e}=\mu_{h}$),
so the region II is absent. For the case of extrinsic graphene with
no photo-inversion ($\mu_{e}=-\mu_{h}$), the phase space for an inter-band
recombination shrinks to zero and the region I disappears, becoming
the case, shown in the Fig. \ref{fig:2.1a}. All other possible
relative values of $\mu_{e}$ and $\mu_{h}$ fall between these boundary
cases.

For comparison, the Drude model, modified to include the inter-band
transitions, has been used in order to calculate the plasmon dispersion
in the case of active and passive doped graphene. These are shown
in the Figs. \ref{fig:2.1a} and \ref{fig:2.1b} as dashed black
curves. As can be seen, they follow the exact plasmon dispersion,
calculated from the non-local quantum model in the regions of small
in-plane wavevectors, but start deviating for the wavevectors, approaching
the region of inter-band transitions.

\section{Analysis of an infinite structure}

\subsection{Bloch Modes}

Let us first consider an infinite periodic structure, consisting
of alternating graphene/dielectric layers and schematically shown in the Fig. \ref{fig:2.2}b. The
plasmons, excited on the outer graphene sheet, are purely evanescent in
the $z-$direction, but they couple to the plasmon resonances of the
subsequent graphene sheets \citep{DasSarma1982}, acquiring a phase lag upon the transition across the conductive layers \citep{A.P.Vinogradov2010,Orlov2014}.
This difference in phase gives rise to a modulation of their amplitude
in space. When this modulation satisfies the periodic Bloch boundary
conditions, it starts tracing an effective wave, propagating down
the structure, with $\operatorname{Re}\left[\text{K}\right]$, being the spatial frequency, and $\operatorname{Im}\left[\text{K}\right]$ - the rate of decay or growth of the wave in space. Due to the convention on forward direction of propagation, adopted in this work, $\operatorname{Im}\left[\text{K}\right]<0$ characterizes waves of spatially decreasing amplitude, and vice versa. The mechanism for an amplification of Bloch waves is based on the coherent amplification of the plasmonic modes due to stimulated electron-hole pair recombination of the photo-inverted graphene sheets, happening in the phase space of the region I in the Fig. \ref{fig:2.1b}.

\begin{figure*}
	\centering
\subfloat[\label{fig:epsilon-x-eff a}]{%
	\begin{minipage}[b][1\totalheight][c]{0.33\textwidth}%
		\includegraphics[scale=0.33]{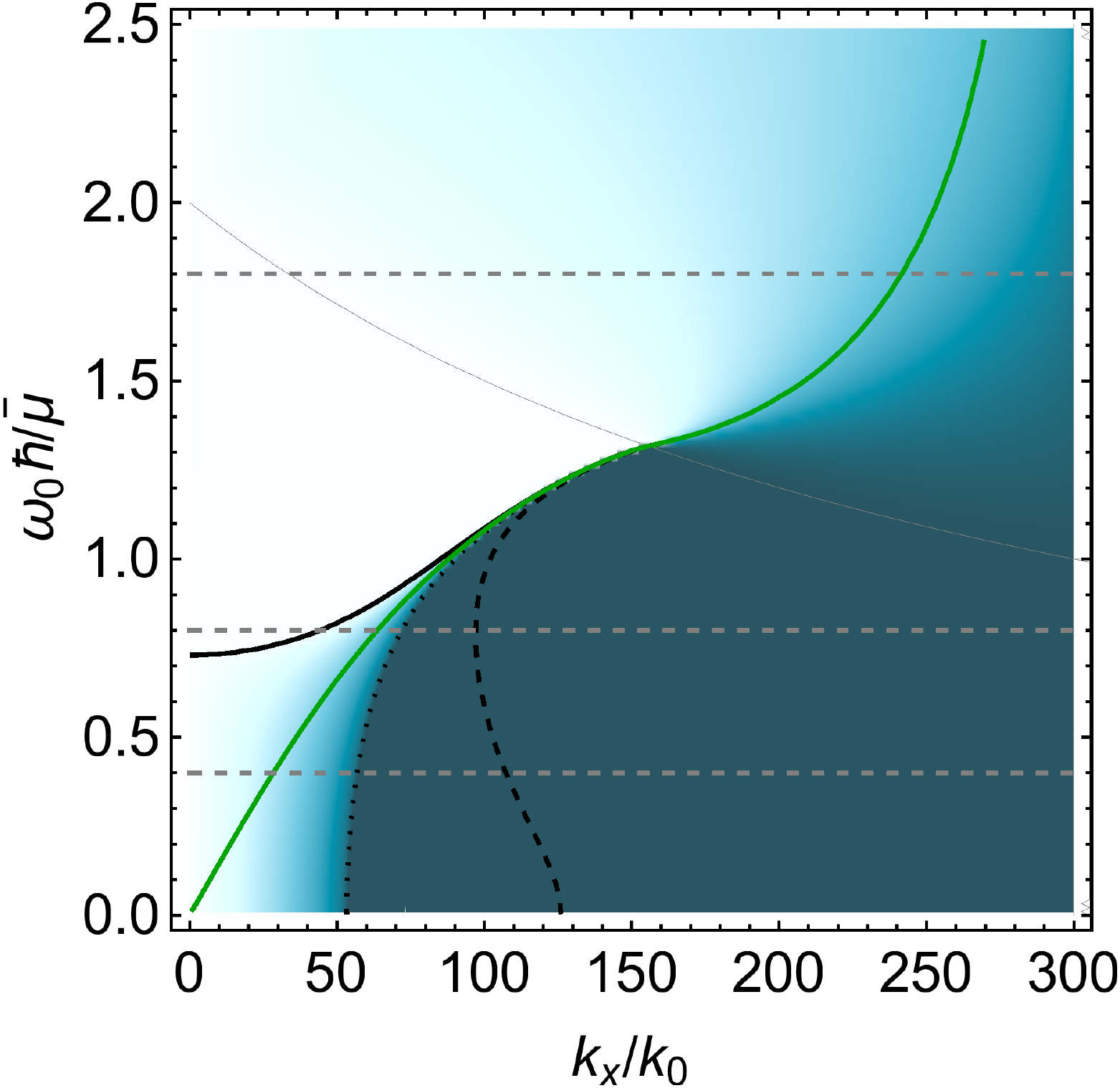}
		\includegraphics[scale=0.33]{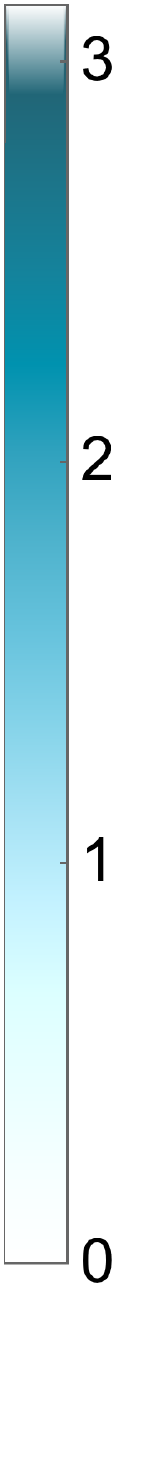}%
	\end{minipage}}
\hspace*{\fill}
\subfloat[\label{fig:epsilon-x-eff b}]{%
\begin{minipage}[b][1\totalheight][c]{0.33\textwidth}%
		\includegraphics[scale=0.32]{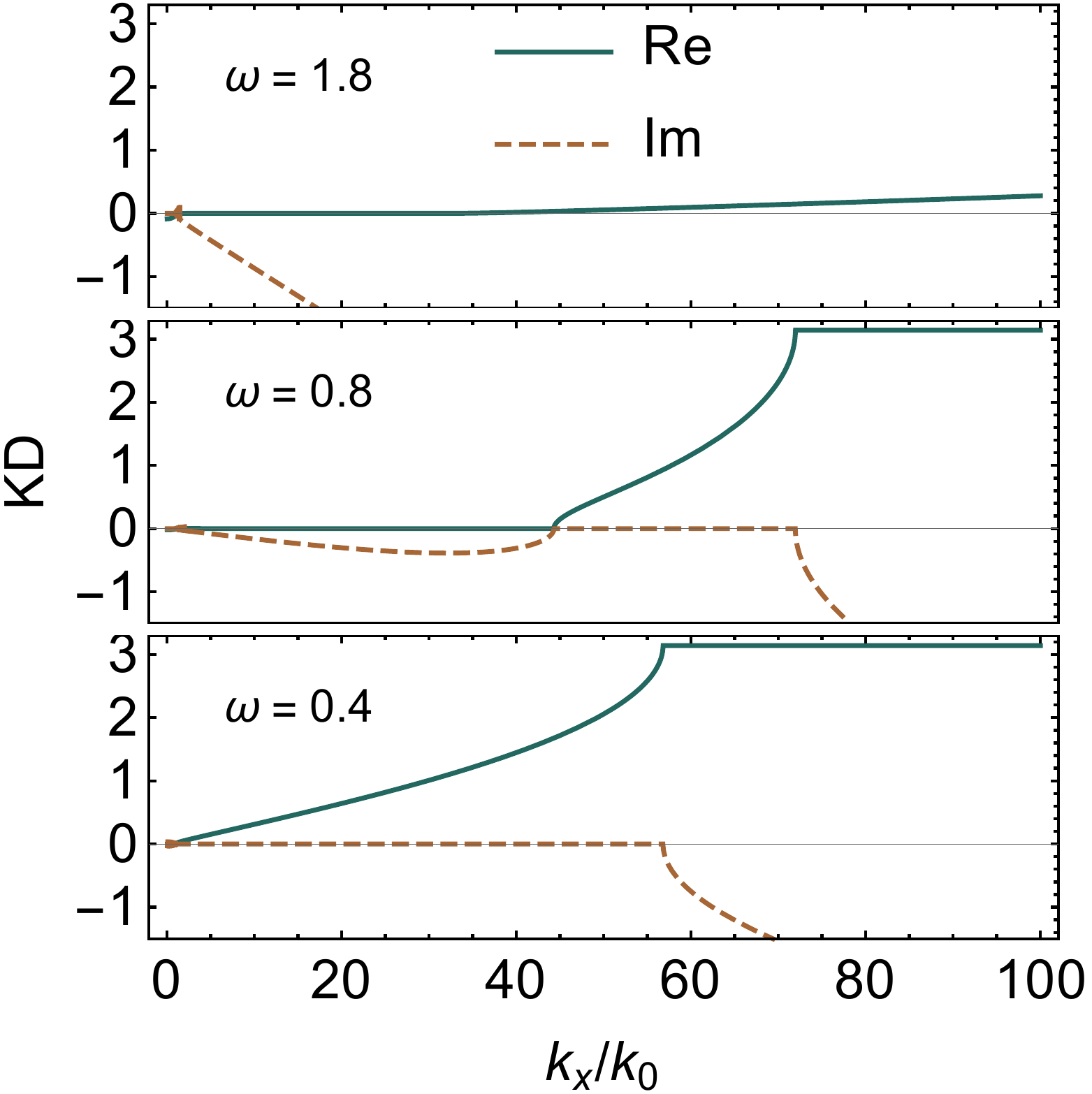}\vspace{-1.5pt}
\end{minipage}
\hspace*{\fill}}
\subfloat[\label{fig:epsilon-x-eff c}]{%
	\begin{minipage}[b][1\totalheight][c]{0.33\textwidth}%
		\includegraphics[scale=0.33]{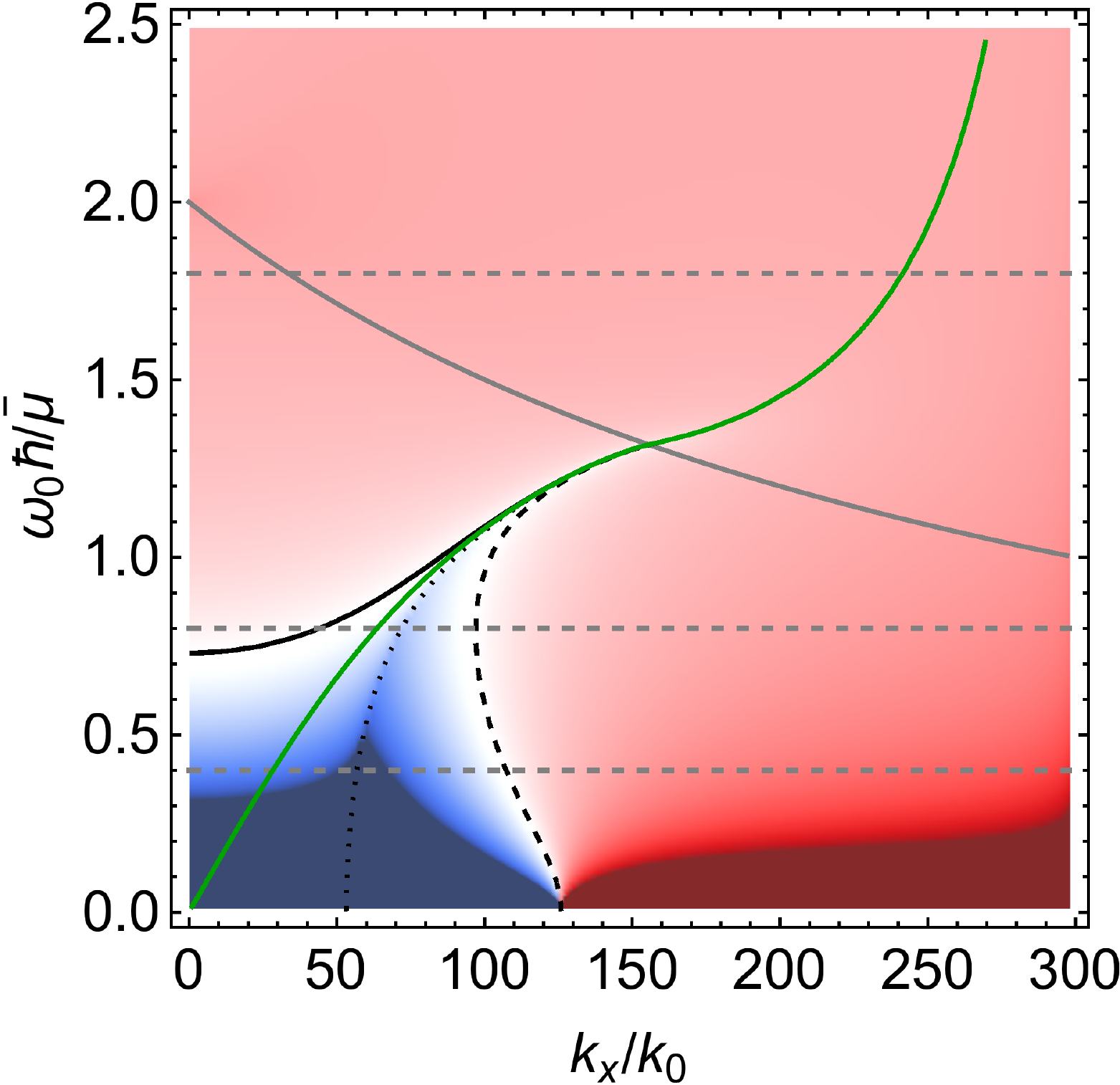}
		\includegraphics[scale=0.33]{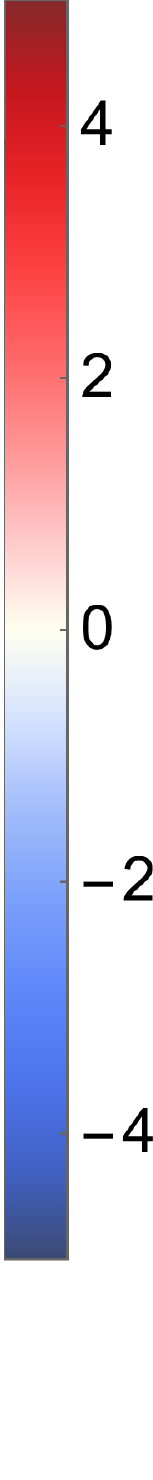}%
	\end{minipage}}

\caption{(a) - real part of the Bloch wavevector $\operatorname{Re}\left[\text{K}D\right]$ for a passive infinite structure is represented as a density map, where solid green line represents the single sheet plasmon dispersion. The black solid and dashed lines bound the region of plasmonic operation of the structure. (b) - the isofrequency contours (horizontal cross-sections of Fig. (a) along the dashed lines) at three different $\tilde{\omega}$ values. (c) - $\operatorname{Re}\left[\varepsilon_{x-eff}\right]$, calculated from the Eq. \eqref{eq:2.5}, is shown as a density plot. Here the condition $\operatorname{Re}\left[\text{K}D\right]=\operatorname{Im}\left[\text{K}D\right]=0$ is marked by the solid black line, the condition $\operatorname{Re}\left[\text{K}D\right]=\operatorname{Im}\left[\text{K}D\right]=\pi$ - by the dashed black line, and the condition $\operatorname{Re}\left[\text{K}D\right]=\pi$
and $\operatorname{Im}\left[\text{K}D\right]=0$ - by the dotted black line. Above calculations were performed for the structures with the following parameters: $\mu_{e}=-\mu_{h}=0.031\,\unit{eV}$, $T = 0\,\unit{eV}$, $\gamma=0$, $\varepsilon_{d}=1$, $d=300\,\unit{nm}$.\label{fig:epsilon-x-eff}}
\end{figure*}

Applying the transfer-matrix method \citep{Yeh2005,Yeh1976} to an infinite
stack of alternating graphene/dielectric layers and imposing periodic
Bloch boundary conditions, the dispersion relation of the Bloch modes
can be established \citep{Iorsh2013,Gu1996,Yeh2005}

\begin{equation}
\cos(\text{K}D)=\cos(\text{k}_{z,1}d_{1})-\frac{iZ_{0}k_{z,1}\sigma_{s}(q,\omega)}{2\varepsilon_{1}k_{0}}\sin(\text{k}_{z,1}d_{1})\label{eq:2.3}
\end{equation}
where $\text{K}$ is the magnitude of the Bloch wave-vector, $D$ - the period of the structure,
$z$ - the axis of anisotropy, $\varepsilon_{1}$ - the permittivity of dielectric,
$d_{1}$ - its thickness, $k_{0}$ - the wavevector of the incident radiation,
$k_{z,1}$ - its $z$-component in the dielectric and $\sigma_{s}(q,\omega)$
- the non-local graphene sheet conductivity. Eq. \eqref{eq:2.3} is
an even function with two generally complex roots of opposite signs:
$\pm\text{K}D$. As an example, $\operatorname{Re}\left[\text{K}D\right]$ is calculated for the structure with doped passive graphene sheets at zero temperature, and presented in Fig. \ref{fig:epsilon-x-eff a}.

For illustration purposes, we consider an infinite structure with the period of $d_{1}=300\,\unit{nm}$ and the permittivity of the dielectric $\varepsilon_{1}=1$. In Fig. \ref{fig:epsilon-x-eff b} we present $\text{K}D$ values (isofrequencies), calculated for three excitation energies $\tilde{\omega}$, marked as dashed lines in Fig. \ref{fig:epsilon-x-eff a}. For $\tilde{\omega}=0.4$, as seen from the bottom panel of Fig. \ref{fig:epsilon-x-eff b}, $\operatorname{Re}\left[\text{K}D\right]=\left[0, \pi\right]$ and $\operatorname{Im}\left[\text{K}D\right]=0$ over the range $k_{x}/k_{0}\in\left[1,\,56\right]$, being characteristic of the hyperbolic operation of the structure (the light cone in this normalized representation is a vertical line at $k_{x}/k_{0}=1$). For these excitation parameters the graphene sheets of the structure are operating in the loss-less region II of the Fig. \ref{fig:2.1}, and thus do not exhibit any inter- or intra-band losses. As a result, the plasmonic modes are loss-less in region $k_{x}/k_{0}\in\left[1,\,56\right]$, as seen from $\operatorname{Im}\left[\text{K}D\right]=0$. For $k_{x}/k_{0}>56$, the structure exhibits loss, as seen from the $\operatorname{Im}\left[\text{K}D\right]<0$. Since excitations occur within the loss-less region II, this damping must originate from the periodic structure itself. The condition $\operatorname{Re}\left[\text{K}D\right]=\abs{\operatorname{Im}\left[\text{K}D\right]}=\pi$ for excitation at the band edge is marked as a dashed black line in Fig. \ref{fig:epsilon-x-eff a}.

In the middle panel of Fig. \ref{fig:epsilon-x-eff b} the loss-less hyperbolic modes are located within $k_{x}/k_{0}\in\left[45,\,97\right]$, exhibiting band gaps for $k_{x}/k_{0}<45$ and $k_{x}/k_{0}>97$, seen as $\operatorname{Im}\left[\text{K}D\right]<0 \cup \operatorname{Re}\left[\text{K}D\right]=0$ and $\abs{\operatorname{Im}\left[\text{K}D\right]}>\pi \cup \operatorname{Re}\left[\text{K}D\right]=\pi$ respectively. The lower boundary of the hyperbolic modes region can be expressed as $\operatorname{Re}\left[\text{K}D\right]=\operatorname{Im}\left[\text{K}D\right]=0$, and is shown as solid black line in Fig. \ref{fig:epsilon-x-eff a}.

In the top panel of the Fig. \ref{fig:epsilon-x-eff b} we can see absence of a hyperbolic mode, and increasingly larger losses, present both in regions II and III (boundary between them shown as a thin gray line in Fig. \ref{fig:epsilon-x-eff a}).

\subsection{Homogenization}

\begin{figure}
\includegraphics[width=1\columnwidth]{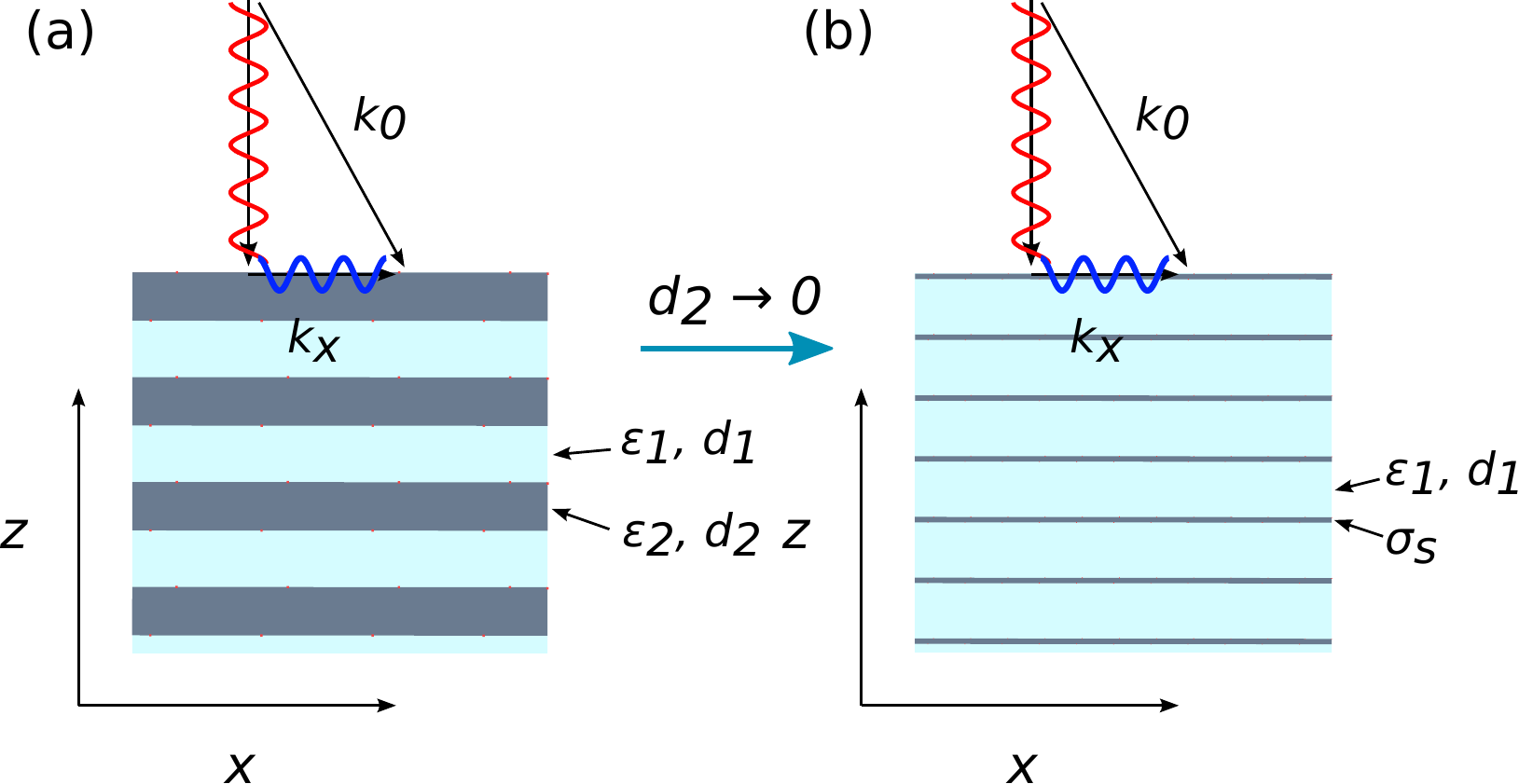}

\caption{In case of the conductive sheets of the finite thickness (a), the incident TM wave couples to carriers oscillations in both $x$- and $z$-directions. In case of infinitesimally thin conductive layers (b), the carriers oscillations are confined to the plain of the layer.\label{fig:2.2}}
\end{figure}

The role of thin metallic layers in a general hyperbolic metamaterial
is to provide a confining medium for plane plasma oscillations, excited by TM polarized excitations (see Fig. \ref{fig:2.2}, here, we are concerned with TM-polarized  waves, as they couple to strongly-bound TM plasmon modes of graphene \citep{Page2015}). The $x-$component of incident wavevectors will couple to the plasma oscillations and at resonant frequencies provide
the necessary negative $\varepsilon_{x-eff}$ component of the dielectric
tensor of the effective medium, giving it overall hyperbolic properties.

In case of a metal with finite thickness, both $\varepsilon_{x-eff}$ and $\varepsilon_{z-eff}$ are affected by the negative $\varepsilon$ of the metal, since due to the metal's finite thickness both in- and out- of plane electric field components couple to the plasma oscillations. A detailed procedure of finding the components of the dielectric tensor of a general layered metamaterial has been described, for example, in \citep{Agranovich1985,Kidwai2012,Wood2006}.

On the contrary, in the case of the metallic layer thickness shrinking to zero, the plasma is confined to two dimensions (Fig. \ref{fig:2.2}b), and so cannot be polarized in $z-$direction. %As a result, the field oscillations, perpendicular to the layers, cannot couple to the carriers, which are confined to the two-dimensional sheets.
The $z-$component of the effective permittivity of the structure is thus described exclusively by the dielectric response: $\varepsilon_{z-eff}=\varepsilon_{d}$. The energy conservation for the effective structure will be expressed as:

\begin{equation}
\frac{k_{x}^{2}}{\varepsilon_{d}}+\frac{\text{K}^{2}}{\varepsilon_{x-eff}}=k_{0}^{2}\label{eq:2.4}
\end{equation}
Here, $\text{K}=\text{K}^{'}+i\text{K}^{''}$ is the wavevector of the Bloch modes of the
structure, calculated from Eq. \eqref{eq:2.3}. From this, the effective
$\varepsilon_{x-eff}$ can be obtained:

\begin{equation}
\varepsilon_{x-eff}=\frac{\text{K}^{'2}-\text{K}^{''2}}{k_{0}^{2}-\frac{k_{x}^{2}}{\varepsilon_{d}}}-2i\frac{\text{K}^{'}\text{K}^{''}}{k_{0}^{2}-\frac{k_{x}^{2}}{\varepsilon_{d}}}\label{eq:2.5}
\end{equation}
where the quantity $k_{0}^{2}-\frac{k_{x}^{2}}{\varepsilon_{d}}$
is negative outside of the light cone. Analysis of Eq. \eqref{eq:2.5} provides us with a tool-box for an analytic calculation of the regions of hyperbolic behavior and the plasmonic band gap of the structure. Concretely, we can find the regions in the $\omega / k_{x}$ phase space where the following conditions are satisfied:

\begin{equation}
\begin{cases}
\operatorname{Re}\left[\text{K}D\right]=0, \, \operatorname{Im}\left[\text{K}D\right]=0 & \left(a\right)\\
\operatorname{Re}\left[\text{K}D\right]=\pi, \, \operatorname{Im}\left[\text{K}D\right]=0 & \left(b\right)\\
\operatorname{Re}\left[\text{K}D\right]=\pi, \, \operatorname{Im}\left[\text{K}D\right]=\pi & \left(c\right) \label{eq:2.6}
\end{cases}
\end{equation}

From Eq. \eqref{eq:2.5} we can see that $\left[\varepsilon_{x-eff}\right]<0$ for $0\leq \text{K}^{''2}\leq \text{K}^{'2}\leq\pi/d_{1}$, with two edge conditions (a) and (c) bounding the phase space of hyperbolic character. The modes are undamped between conditions (a) and (b), and experience spatial damping between conditions (b) and (c). We can derive a corresponding set of analytic expressions of the boundaries in the $\omega / k_{x}$ space, on which the conditions \eqref{eq:2.6} are satisfied. In terms of the imaginary part of graphene's non-local conductivity $\sigma_{s}^{''}$ for the loss-less case these boundaries have the following form(details of the derivation may be found in the appendix):

\begin{equation}
\begin{cases}
\sigma_{s}^{''}(q,\omega)=\frac{2\varepsilon_{1}}{Z_{0}a}\tanh\left(\frac{k_{0}d_{1}a}{2}\right) & \left(a\right)\\
\sigma_{s}^{''}(q,\omega)=\frac{2\varepsilon_{1}}{Z_{0}a}\left(\coth\left(\frac{k_{0}d_{1}a}{2}\right)+\frac{1}{\sinh\left(\frac{k_{0}d_{1}a}{2}\right)}\right) & \left(b\right)\\
\sigma_{s}^{''}(q,\omega)=\frac{2\varepsilon_{1}}{Z_{0}a}\frac{\cosh\left(k_{0}d_{1}a\right)+\cosh\left(\pi\right)}{\sinh\left(k_{0}d_{1}a\right)} & \left(c\right)
\end{cases}\label{eq:2.7}
\end{equation}

These conditions are satisfied along the solid, dotted and dashed black lines respectively, as presented in Figs. \ref{fig:epsilon-x-eff a} and \ref{fig:epsilon-x-eff c}. The black lines in Fig. \ref{fig:epsilon-x-eff c} converge to the single sheet plasmon dispersion for large values of the excitation frequency and its in-plane component. Increasing thickness of the dielectric layer weakens the coupling of the plasmonic modes between the graphene sheets, and thus the region of $\operatorname{Re}\left[\varepsilon_{x-eff}\right]<0$ would shrink. These and other properties can be inferred from an analysis of Eqs. \eqref{eq:2.7}, which are central to the study of the hyperbolic behavior of the considered plasmonic metamaterials.

\subsection{Mode amplification in active HMMs}

\begin{figure*}
	\centering
\subfloat[\label{fig: Im=00005BKD=00005D-a}]{%
	\begin{minipage}[b][1\totalheight][c]{0.49\textwidth}%
		\includegraphics[scale=0.5]{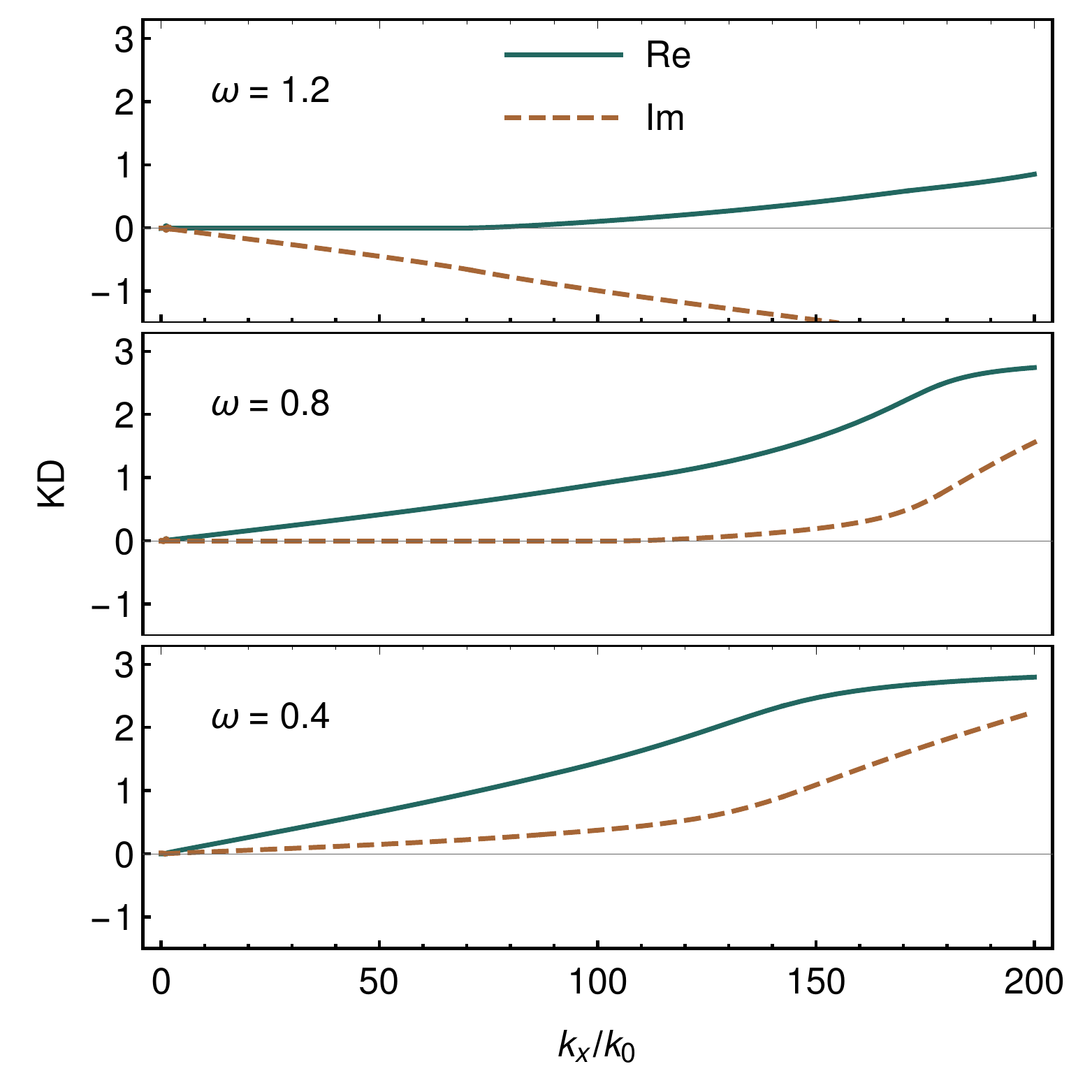}\vspace{-1.5pt}
	\end{minipage}
}%\hspace*{\fill}
\subfloat[\label{fig: Im=00005BKD=00005D-b}]{%
	\begin{minipage}[b][1\totalheight][c]{0.49\textwidth}%
		\includegraphics[scale=0.49]{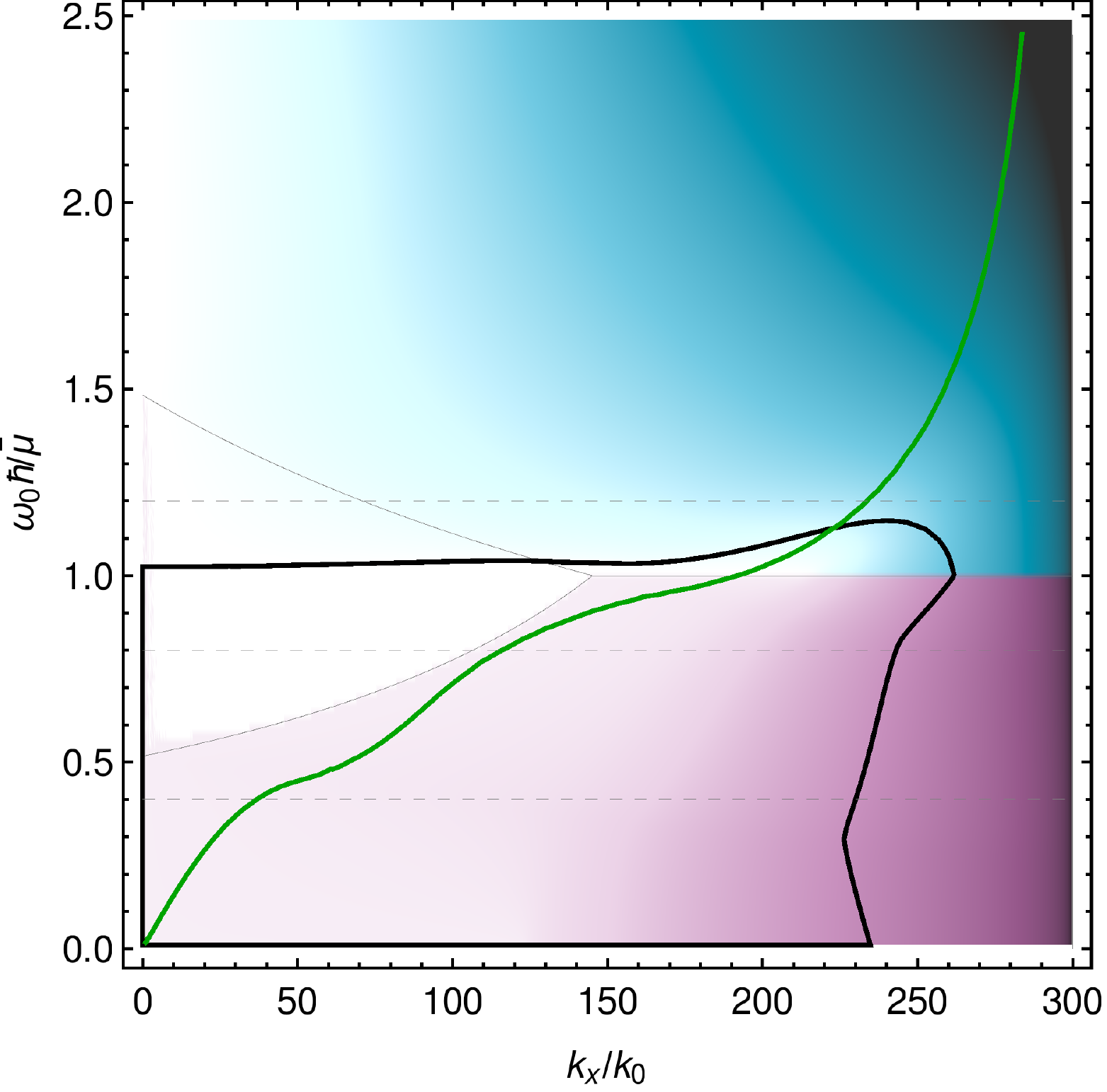}\,
		\includegraphics[scale=0.41]{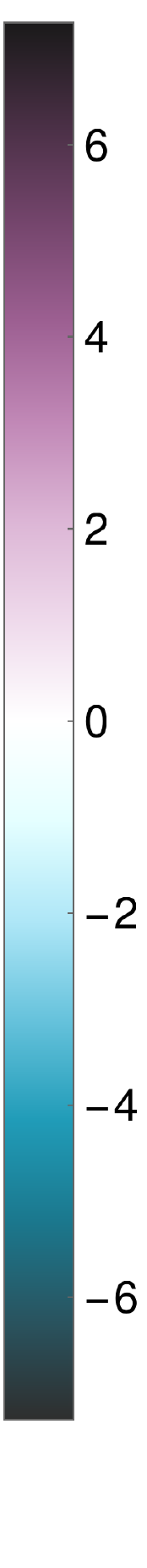}%
	\end{minipage}
}

\caption{(a) - real (teal) and imaginary (brown) parts of the isofrequency curves are plotted for various values of $\omega_{0}\hbar/\mu$. (b) - the imaginary part of the Bloch wavevector is shown for the HMM, based on doped
graphene with photo-excitation. Thin gray lines mark the boundaries of regions of behavior, corresponding to those in Fig. \ref{fig:2.1b}. The black
curve bounds the region of $\varepsilon_{x-eff}\leq0$. Dashed gray lines correspond to the
cross-sections in $\omega_{0}\hbar/\mu$, for which the isofrequency
curves are shown in (a). For reference, the green line is the frequency
dispersion of the single graphene sheet, suspended in a dielectric with $\varepsilon_{d}$. Parameters of the structure: $\mu_{e}=0.023\,\unit{eV}$, $\mu_{h}=0.008\,\unit{eV}$, $\varepsilon_{d}=1$, $d=50\,\unit{nm}$.
\label{fig: Im=00005BKD=00005D1-1}}
\end{figure*}

Plasmonic metamaterials are inherently lossy due to their exploitation of the plasma resonances. This causes the wavevector of the plasmonic modes to have a large imaginary part, corresponding to a decay of the modes in space. In order to overcome this problem, it was proposed to incorporate gain medium in the dielectric layers \citep{Soukoulis2010,Xiao2010,Savelev2013}. Here, we propose to explore gain, available from graphene by optically pumping the carrier plasma into a state of inversion. Ideally, the plasmons, coupling to the inverted carriers, would cause their stimulated recombination and undergo amplification. Here we employ the exact in RPA quantum conductivity model in order to describe the response of the inverted carrier plasma
in the doped graphene sheets.

In the case of a single graphene sheet, the conditions for excitation
of plasmons are satisfied on the line in the $\left(q,\,\omega\right)$
space (Figs. \ref{fig:2.1a} and \ref{fig:2.1b}), while in the
case of the multiple coupled graphene sheets, the whole region of
plasmonic modes opens up in the $\left(q,\,\omega\right)$ space,
as seen from the $\operatorname{Re}\left[\varepsilon_{x-eff}\right]<0$ region in Fig. \ref{fig:epsilon-x-eff a}.

A non-zero imaginary part of the wavevector defines the change in the
amplitude of the Bloch wave, as it propagates through the structure.
This parameter acquires meaning only in conjunction with $\operatorname{Re}\left[\text{K}D\right]$,
such that $\operatorname{Im}\left[\text{K}D\right]<0\,\cup\,\left|\operatorname{Im}\left[\text{K}D\right]\right|<\left|\operatorname{Re}\left[\text{K}D\right]\right|$
corresponds to damped modes, $0<\operatorname{Im}\left[\text{K}D\right]<\operatorname{Re}\left[\text{K}D\right]$ -
to amplified modes and $\left|\operatorname{Im}\left[\text{K}D\right]\right|>\operatorname{Re}\left[\text{K}D\right]$ - to
unstable modes. Therefore, generally, stable modes have to satisfy

\begin{equation}
\left|\operatorname{Im}\left[\text{K}D\right]\right|\leq\left|\operatorname{Re}\left[\text{K}D\right]\right|\label{eq:3.1}
\end{equation}

To illustrate this, we are using the quantum conductivity model
together with Eq. \eqref{eq:2.3} to perform a calculation of $\operatorname{Im}\left[\text{K}D\right]$ for an active structure, which is plotted in Fig. \ref{fig: Im=00005BKD=00005D-b}. The regions of gain (purple) and loss (blue) of the
considered GHMM coincide with the single graphene sheet regions of
the inter-band recombination and excitation respectively. The condition $\operatorname{Re}\left[\varepsilon_{x-eff}\right]<0$ is satisfied over the region which is bound by the black curve in the Fig. \ref{fig: Im=00005BKD=00005D-b}. As can be seen, its location is independent of the zones of the inter/intra-band processes in the single-sheet graphene, and spans over both amplifying and lossy regions.

The region where $\operatorname{Re}\left[\varepsilon_{x-eff}\right]<0$ in Fig. \ref{fig: Im=00005BKD=00005D-b} is shown for a particular thickness of the dielectric layer. Varying the dielectric thickness changes the coupling strength of the plasmonic modes \citep{DasSarma1982}, thus changing the phase space of the plasmonic region, with the top right corner following the single sheet plasmon dispersion, and approaching the origin with an increase of the dielectric thickness.

The isofrequency contours of representative modes are shown in Fig. \ref{fig: Im=00005BKD=00005D-a} for three different values of the excitation frequency $\omega_{0}\hbar/\mu$ as real and imaginary parts of the isofrequency curves. For the frequency $\omega_{0}\hbar/\mu=0.4$ the hyperbolic mode is seen as a continuous curve $\operatorname{Re}\left[\text{K}D\right]\in\left[1,\,\pi\right)$ for the in-plane wavevectors of $k_{x}/k_{0}\in\left[1,\,200\right]$. It is amplified in this range, but amplification is stable only for $k_{x}/k_{0}\in\left[1,\,230\right]$, - within the $\epsilon_{x-eff}\leq0$ loop in Fig. \ref{fig: Im=00005BKD=00005D-b}.
Analogous behavior is observed for excitation frequencies in the
range $\omega_{0}\hbar/\mu\in\left(0,\,0.5\right)$, where the in-plane
components are fully within region I.

For $\omega_{0}\hbar/\mu=0.8$, the modes for $k_{x}/k_{0}\in\left[1,\,110\right]$ are located within the loss-less region II,  where $\operatorname{Im}\left[\text{K}D\right]\approx0$ and $\operatorname{Re}\left[\text{K}D\right]>0$. For $k_{x}/k_{0}\in\left[110,\,240\right]$ the modes enter the region of gain I, thus being amplified, with $\operatorname{Im}\left[\text{K}D\right]>0$.

The last set of modes is demonstratively represented at $\omega_{0}\hbar/\mu=1.2$. There is a band-gap within region II for $k_{x}/k_{0}\in\left[1,\,80\right]$, where $\operatorname{Re}\left[\text{K}D\right]\approx0$ and $\operatorname{Im}\left[\text{K}D\right]<0$. The over-damped mode exists for $k_{x}/k_{0}>80$ due to the inter-band loss of region III, preventing perfect destructive interference, which would be causing the band-gap.

\section{Pulse propagation in a finite stack}

\subsection{Loss Compensation}

So far we have considered light propagating in infinite periodic stacks. Despite being a good approximation for a description of sufficiently large structures, it is not clear how this model would generalize to structures at the current level manufacturing possibilities, consisting of just several layers. To answer this question we use the transfer matrix method (TMM), which allows us to spatially resolve the electric and magnetic fields, as well as calculate the Poynting vector, inside of the finite structure with the given set of input or output fields. With this we calculate the quasi-static field distribution, resulting from the illumination of the finite stack with a narrow spatial Gaussian intensity envelope. The Gaussian envelope is sufficiently small in real space, thus consisting of $k_{x}$ components that are evanescent in free space, and coupling to the plasmonic modes of the considered heterostructure.

\begin{figure*}
\subfloat[\label{fig: Pulse, T=00003D0, gamma=00003D0, Passive}]{\includegraphics[scale=0.47]{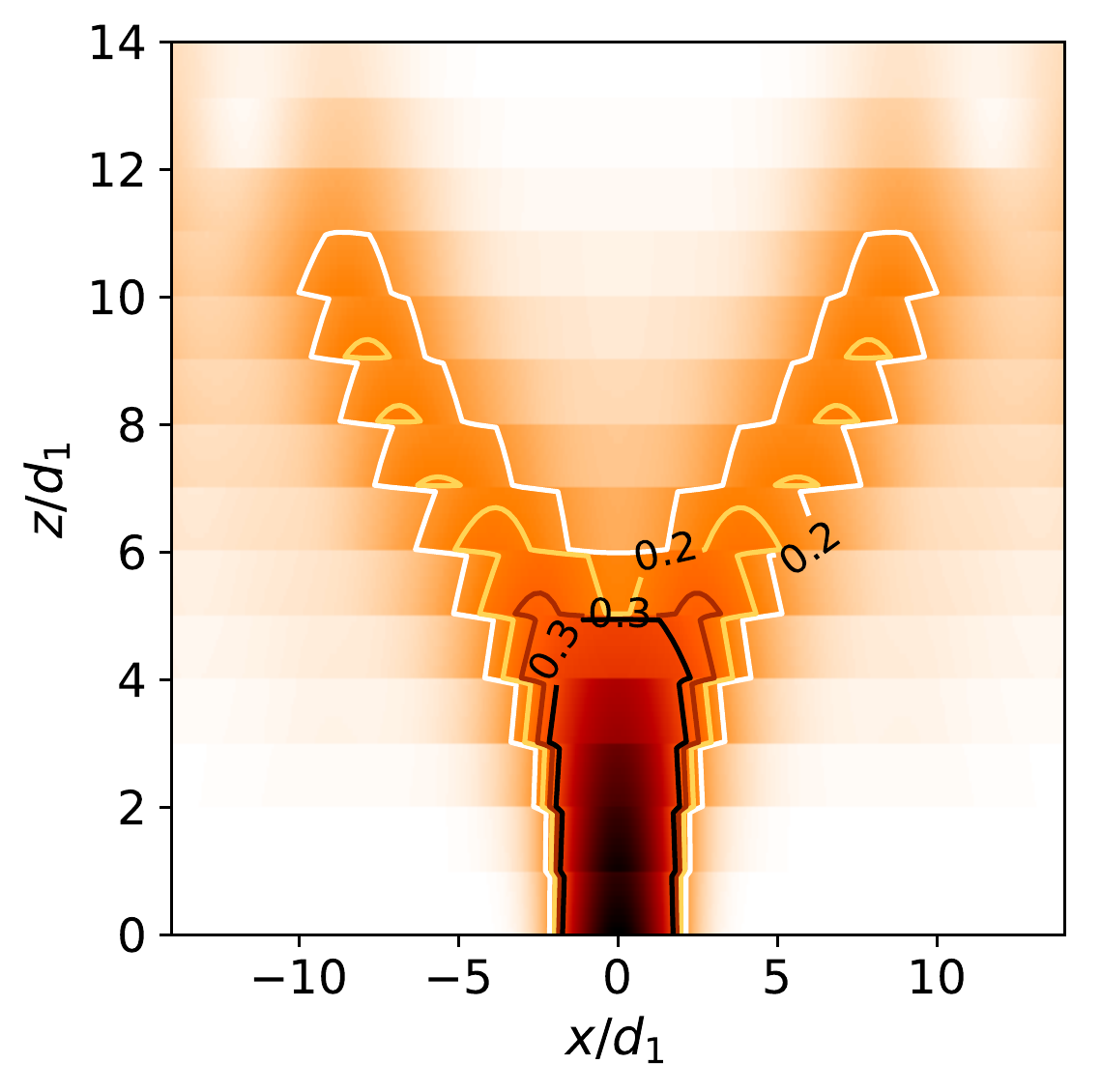}

}\subfloat[\label{fig: fig: Pulse, T>0, gamma>0, Passive}]{\includegraphics[scale=0.47]{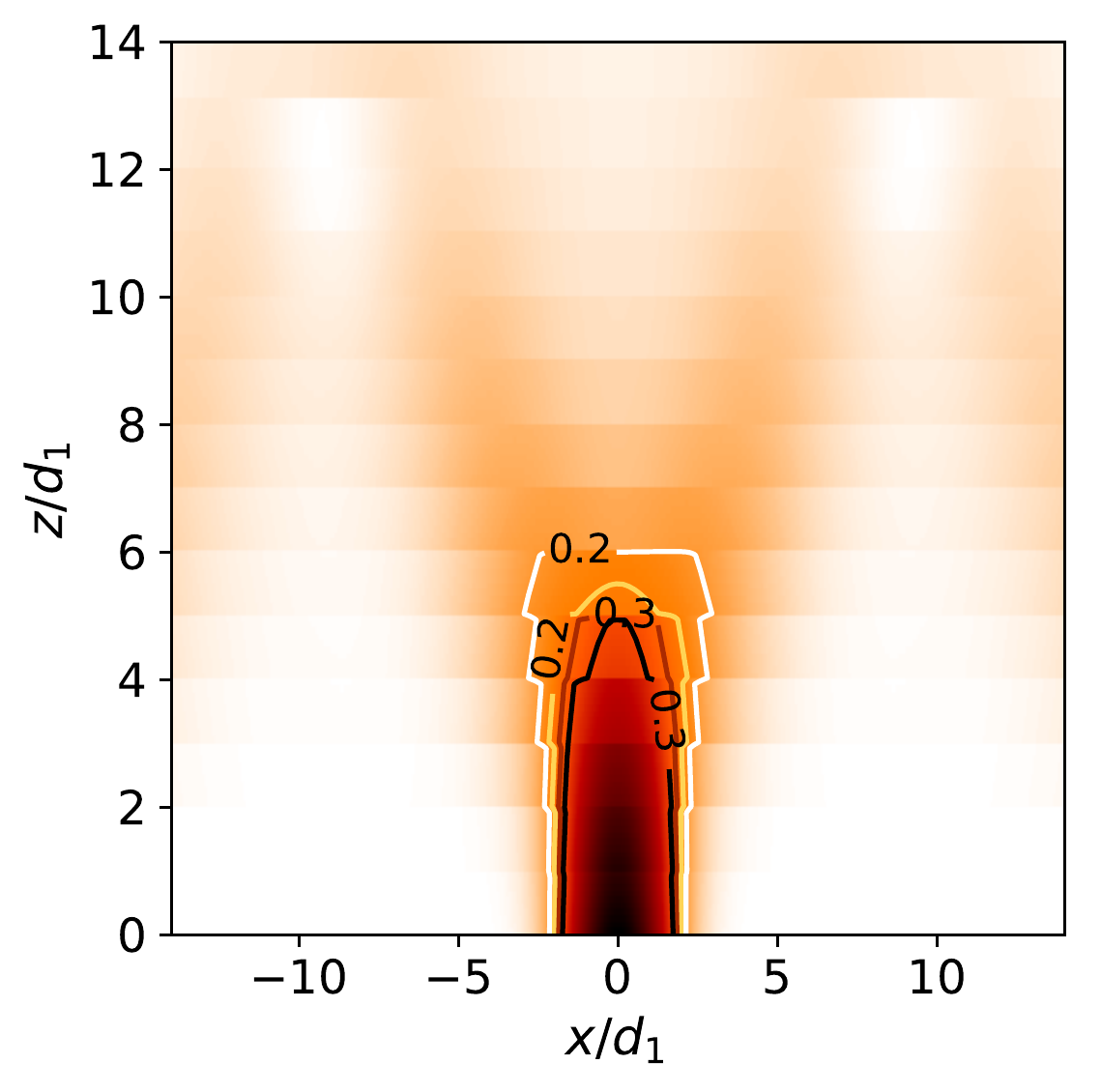}

}\subfloat[\label{fig: fig: Pulse, T>0, gamma>0, Active}]{\includegraphics[scale=0.47]{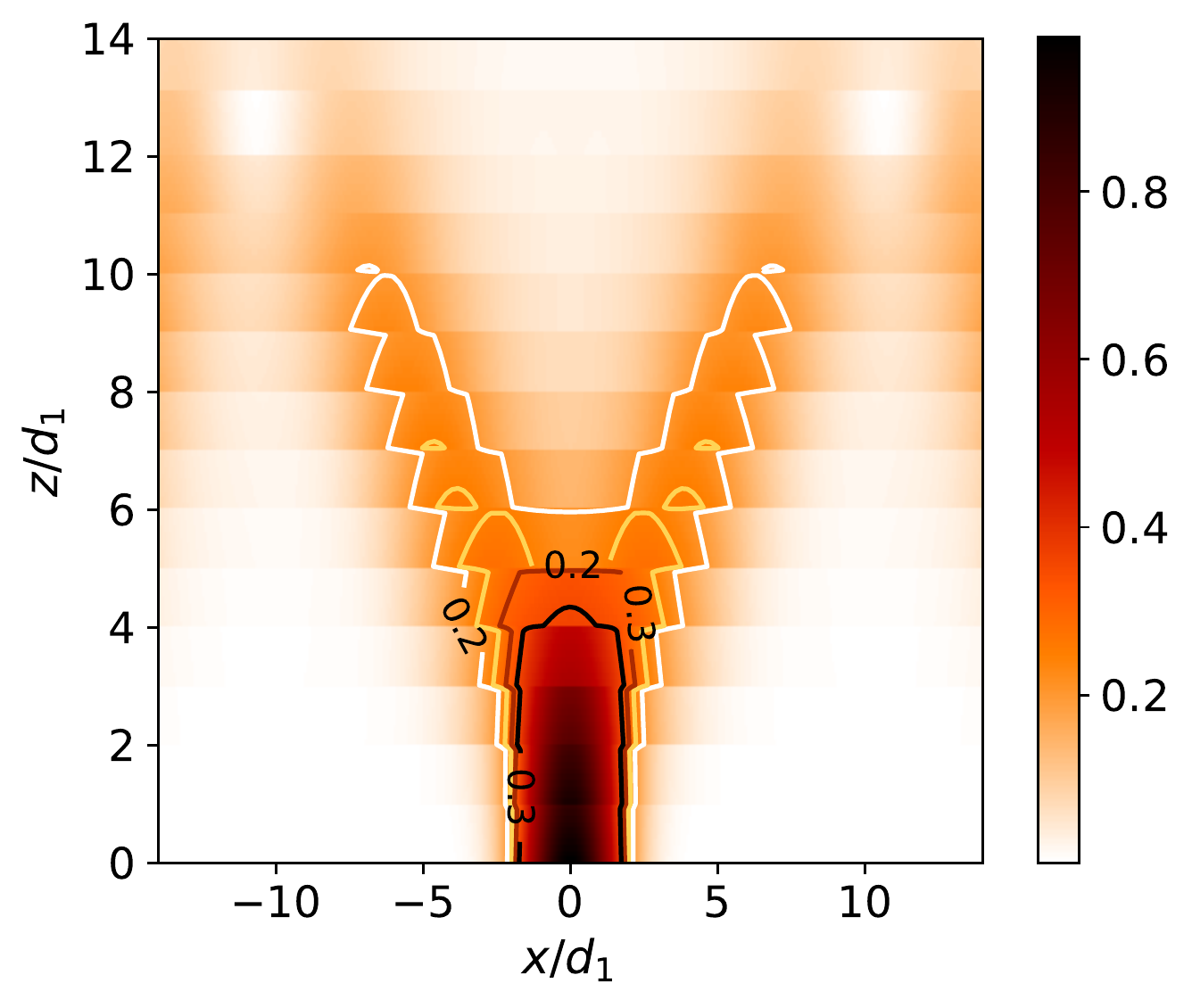}

}\caption{$\abs{E_x}^2$ field distribution, corresponding to the propagation of a Gaussian envelope through the finite graphene-dielectric stack, composed of 14 layers with $d_1=50\,\unit{nm}$, is shown as density maps. Figs. (a) and (b) are calculated for the passive case, with the graphene sheets doped to $\mu_e = -\mu_h = 0.031\,\unit{eV}$. For  Figs. (a) and (b) $T=0$, $\gamma \rightarrow 0$, and $T=0.017\,\unit{eV}$, $\gamma = 0.015\,\unit{eV}$ respectively. Fig. (c) shows the active case with the same losses as in Fig. (b): $T=0.017\,\unit{eV}$, $\gamma = 0.015\,\unit{eV}$, and doping $\mu_e = 0.023\,\unit{eV}$, $\mu_h = 0.008\,\unit{eV}$.
\label{fig: Pulse in finite structure}}
\end{figure*}

We start with calculations of the forward-traveling component of the electric
field distribution in a structure consisting of 14 unit cells, as shown in the Fig. \ref{fig: Pulse, T=00003D0, gamma=00003D0, Passive}. In this initial scenario the graphene sheets are assumed to be at zero temperature, have a small amount of scattering loss to ensure numerical stability of the calculations, and doped at a level higher than the energy of the exciting filled. Thus the structure is operating in the loss-less region II in Fig. \ref{fig:2.1a}. Therefore, the propagating fields do not undergo any damping or loss, though the overall intensity is decreasing due to the fields being reflected from the subsequent graphene-dielectric boundaries (this can also be seen as increase of intensity of the backward-propagating fields).
% Due to absence of loss, there is no energy propagation into or out of the structure, which is confirmed by the calculations of Poynting vector, resulting in $S_{z}=0$ everywhere in the structure.

Subsequently, we keep the parameters of the structure and
stimulation the same, but introduce a finite temperature (resulting in a smearing
of the graphene Fermi level, while graphene is still doped to the same level),
and add collision loss in the graphene sheets. The calculated field intensity
distribution is shown in Fig. \ref{fig: fig: Pulse, T>0, gamma>0, Passive}.
As a result, the depth of propagation of the pulse is greatly reduced
as compared to the loss-less case. The propagating pulse also undergoes
dispersion, as can be seen from the increased width of the pulse.
We notice that the presence of loss introduces an energy flow within the structure, manifested in $S_{z}>0$. This results in degraded imaging performance
of the stack.

Finally, we introduce gain to our heterostructure in the form of optical
inversion of graphene, entering the model as carrier imbalance of
the electrons and holes. The temperature conditions and scattering losses
are yet again the same as in the previous calculation. The resultant field distribution is presented in Fig. \ref{fig: fig: Pulse, T>0, gamma>0, Active}.
It can be seen that the length of propagation of the pulse within
the structure has increased as compared with the scenario in the Fig.
\ref{fig: fig: Pulse, T>0, gamma>0, Passive}, as well as showing
decreased dispersion (the pulse branches width is decreased as compared
to the Fig. \ref{fig: fig: Pulse, T>0, gamma>0, Passive}). The
energy lost due to the temperature effects and scattering is now well
compensated. %, resulting in $S_{z}\rightarrow0$ everywhere in the structure.
All this demonstrates that with added gain in the form of optical inversion
of graphene we can greatly improve pulse propagating properties of
the finite stack, even ensuring stable operation with $S_{z}\rightarrow0$.

\subsection{Using results for infinite structures for the analysis of finite structures}

So far we have computationally investigated the propagation of a small Gaussian envelope in a finite stack, as well as studied the behavior of an infinite
stack, deriving a set of analytic tools for analysis and prediction
of the properties of an infinite structure. An interesting question
to answer is would it be possible to use analytic approach used for an
infinite structure in order to predict behavior of a finite stack,
without having to perform computationally intense simulations of the
finite stack?

Let us consider the TMM framework (which provides the basis for the analysis of a finite stack). To propagate a narrow spatial envelope through the stack we decompose it into a finite number of Gaussian spatial components $k'_{x}$, propagating them individually through the stack and recombining them at the output end with their Fourier weights. In doing so we impose a condition on
the fields at the output layer to be traveling out of the structure
only. This means that the obtained field distributions for a particular
$k'_{x}$ in general are not the eigenmodes of the structure,
but are a linear combination of two counter-propagating eigenmodes.
The eigenmodes of the finite structure are calculated as eigenmodes
of its single unit cell (that is also what we are calculating
in Eq. \eqref{eq:2.3}).

Since the eigenmodes of the unit cell of infinite and finite structures
are calculated identically, we can use the isofrequency contours of
an infinite structure to predict the behavior of a finite one. 
Therefore, in order to ensure a loss-less propagation of the pulse
through the finite structure, the Fourier components of the pulse
should be located within the range of $k_{x}/k_{0}$ where $\operatorname{Im}\left[\text{K}D\right]=0$. At the same
time, regimes where $\operatorname{Im}\left[\text{K}D\right]>0$, although showing gain, are also detrimental
to the pulse propagation, resulting in non-zero Poynting vector. In this regard, $k_{x}/k_{0}$ can be treated as a resolution limit of a finite structure, as it defines the minimum size of the pulse which can propagate, and can be obtained from the isofrequency contours for an infinite structure for the condition where $\operatorname{Im}\left[\text{K}D\right]=0$.

As was shown earlier, the condition of $\operatorname{Im}\left[\text{K}D\right]=0$ is satisfied when
the stack is supporting plasmonic modes in the loss-less case, or
in case of losses compensated by gain. Therefore, the resolution limit
of the structure is defined as the boundary of plasmonic band of operation $\varepsilon_{x-eff}=0$, shown as black contours in Figs. \ref{fig:epsilon-x-eff} and \ref{fig: Im=00005BKD=00005D1-1}. The analytic calculation of this condition
is described in the Section III and in the Appendix.

\section{Conclusion}

We have studied the behavior of graphene-based hyperbolic metamaterials. Employing an RPA model for the graphene conductivity, which accounts for the
plasmonic gain and loss, we developed an analytic tool-set for the characterization of the regions of behavior of the infinite structures. This allows us to uncover the possibility of plasmonic amplification in optically excited heterostructures. Considering structures with the finite number of layers, we demonstrate that the theory assuming infinite structures has a good applicability there as well. We also demonstrated the quasi-equilibrium field distributions of the Gaussian envelopes propagating in graphene-based hyperbolic metamaterials, and showed that these structures can be used to compensate occurring graphene losses.

\section{Appendix}

\begin{figure*}
\subfloat[\label{fig:A1a}]{\includegraphics[width=0.5\linewidth]{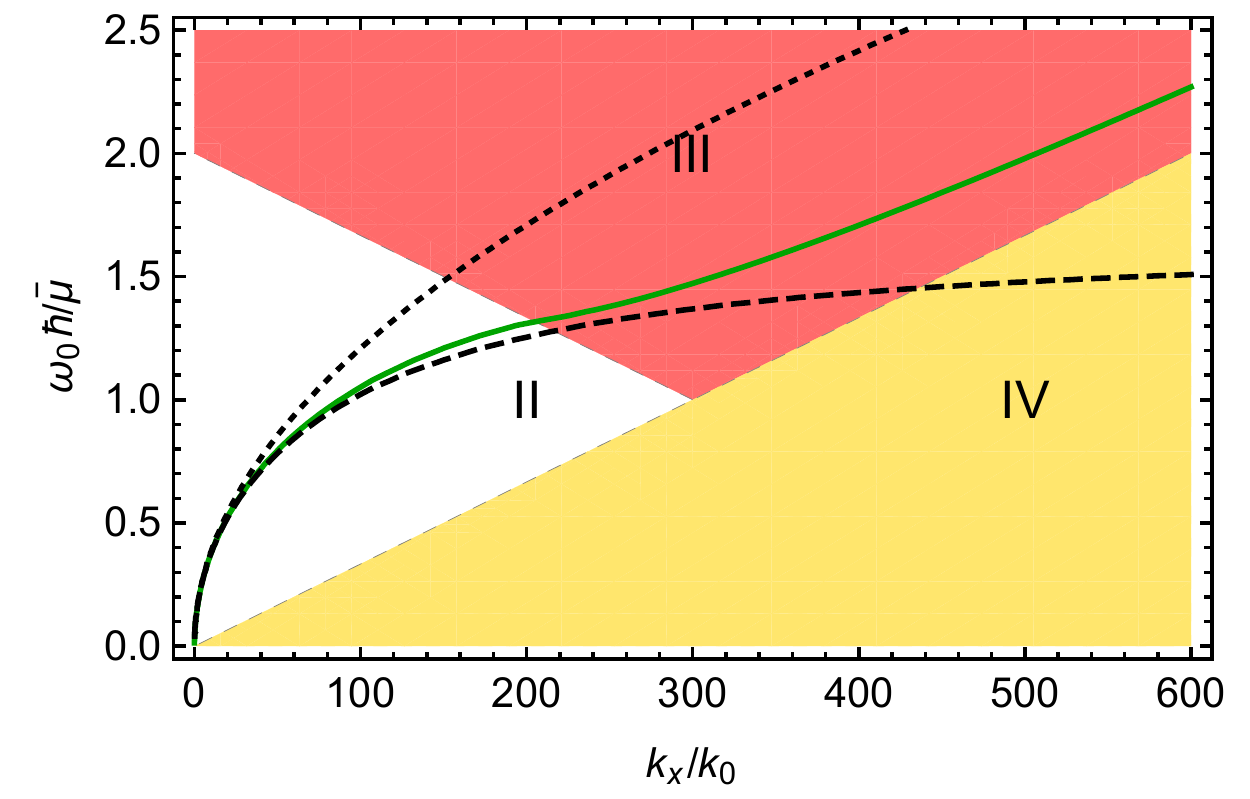}}\hfill{}\subfloat[\label{fig:A1b}]{\includegraphics[width=0.5\linewidth]{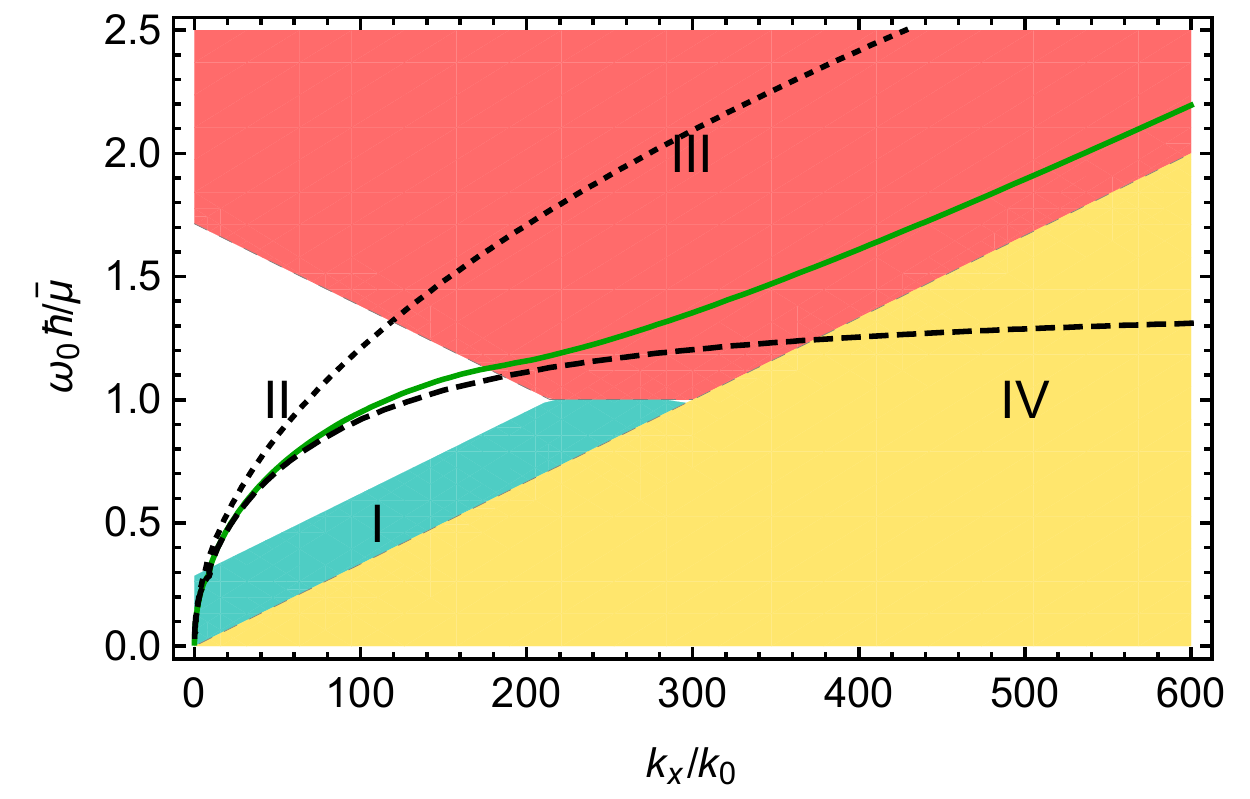}}

\caption{Plasmon frequency dispersion of an air-suspended single sheet of doped
graphene (a) without photo-inversion and (b) with photo-inversion.
The non-local quantum conductivity model was used to obtain solid
green curves, and modified local Drude model was used to obtain dashed
curves. The vertical axis is scaled with $\omega_{pl}\hbar/\bar{\mu}$.
Plasmons experience gain in the region I, loss in the regions III
and IV due to the inter- and intra-band excitations respectively, and
no loss or gain within the region II.\label{fig:A1}}
\end{figure*}

\subsection{Quantum polarizability model}

In general, $\Pi(q,\omega)$ in the Eq. \eqref{eq:2.1-1} is the irreducible
polarizability, quantifying the response of the particle/hole plasma
to the external excitations. In RPA, it is approximated by its first
order term, describing the non-interacting particle hole plasma, and
is expressed via the Lindhard Formula:

\begin{equation}
\Pi[n](q,\omega)=\frac{g}{A}\underset{s,s'=\pm}{\sum}\underset{\mathbf{k}}{\sum}\frac{M_{\mathbf{k},\mathbf{k}+\mathbf{q}}^{ss'}[n(\epsilon_{\mathbf{k}}^{s})-n(\epsilon_{\mathbf{k}+\mathbf{q}}^{s'})]}{\epsilon_{\mathbf{k}}^{s}-\epsilon_{\mathbf{k}+\mathbf{q}}^{s'}+\hbar\omega+i\times0}\label{eq:A.1}
\end{equation}

This expression sums over all possible intra-band $\left(\mathbf{k}\right)\rightarrow\left(\mathbf{k}+\mathbf{q}\right)$
and inter-band $s\rightarrow s^{'}$ transitions, contains spin/valley
degeneracy within the constant $g=4$, and is weighted with the square
of the transition matrix $M_{\mathbf{k},\mathbf{k}'}^{ss'}=[1+ss'\cos(\theta_{k,k'})]/2$.

From Eq. \eqref{eq:A.1} an analytic expression for polarizability of
gap-less graphene at zero temperature has been derived in \citep{Page2015,Pyatkovskiy2009a},
and with some rescaling can be expressed as \citep{Page2016}: 

\begin{equation}
\Pi\left(q,\omega\right)|_{\mu}^{T=0}=\frac{g\mu}{8\pi\hbar^{2}\nu_{F}^{2}}\tilde{\Pi}\left(\frac{\hbar\nu_{F}q}{\mu},\frac{\hbar\omega}{\mu}\right)\label{eq:A.2}
\end{equation}
where $\tilde{\Pi}\left(\tilde{q},\tilde{\omega}\right)=-4+\tilde{q}^{2}\frac{G^{+}\left(\frac{2+\tilde{\omega}}{\tilde{q}}\right)+G^{-}\left(\frac{2-\tilde{\omega}}{\tilde{q}}\right)}{2\sqrt{\tilde{q}^{2}-\tilde{\omega}^{2}}}$
and $G^{\pm}\left(z\right)=z\sqrt{1-z^{2}}\pm i\operatorname{arccosh}\left(z\right)$.
This expression describes the polarizability of doped graphene sheet
at equilibrium, with the corresponding chemical potential $\mu$.
The particle-hole symmetry is implicit in $\Pi\left(q,\omega\right)|_{\mu}^{T=0}=\Pi\left(q,\omega\right)|_{-\mu}^{T=0}$.
The expression \eqref{eq:A.2} is analytic for $\operatorname{Im}\left[\omega\right]>0$,
and can be evaluated for $\operatorname{Im}\left[\omega\right]<0$ via the analytic
continuation \citep{Page2015}.

The single sheet plasmon dispersion is calculated exactly for the
doped equilibrium graphene and is shown in the Figs. \ref{fig:2.1a}
and \ref{fig:A1a} as a green solid line. It begins in the region
II, where the excitation energies are below the chemical potential
of doping, thus are insufficient to decay into the inter-band transitions
($\gamma_{pl}=0$). At zero doping region II disappears, and all phase
space within the Dirac cone is occupied by the region III. After region
II the plasmon dispersion passes through the region III with Landau
damping; region IV with intra-band excitations and loss-free region
V.

Optical pumping of graphene would create a bath of excited electron-hole
pairs. This opens up a region I in the phase space of regions II and
III. The plasmons within region I would trigger recombination of electron-hole
pairs, thus undergoing coherent amplification. For comparison, the
single-sheet plasmon dispersion of the extrinsic active graphene is
shown in the Fig. \ref{fig:A1b}. 

\subsection{Modes of passive graphene-based HMM}

The sheet conductivity $\sigma_{s}(q,\omega)$ in the Eq. \eqref{eq:2.3}
is a complex quantity, and can be expressed as $\sigma_{s}(q,\omega)=\sigma_{s}^{'}+i\sigma_{s}^{''}$.
Then, \eqref{eq:2.3} becomes:

\begin{align}
\cos\left(\text{K}D\right) & =\cosh\left(k_{0}d_{1}a\right)-\frac{Z_{0}a\sigma_{s}^{''}}{2\varepsilon_{1}}\sinh\left(k_{0}d_{1}a\right)\nonumber \\
 & +\frac{iZ_{0}a\sigma_{s}^{'}}{2\varepsilon_{1}}\sinh\left(k_{0}d_{1}a\right)\label{eq:B.1}
\end{align}

were $a=-i\mbox{\ensuremath{\sqrt{-\left(k_{x}/k_{0}\right)^{2}+\varepsilon_{d}}}}$.
The Eq. \eqref{eq:A.2} yields $\sigma_{s}^{'}=0$ in the region II
in the Figs. \ref{fig:2.1} and \ref{fig:A1a}, as calculated for
the passive graphene at $T=0$. Thus, Eq. \eqref{eq:B.1} becomes:

\begin{equation}
\cos\left(\text{K}D\right)=\cosh\left(k_{0}d_{1}a\right)-\frac{Z_{0}a\sigma_{s}^{''}}{2\varepsilon_{1}}\sinh\left(k_{0}d_{1}a\right)\label{eq:B.2}
\end{equation}

Now it is possible to establish the analytic solution for boundaries
of stable hyperbolic behavior of the structure, with $\varepsilon_{x-eff}<0$.
The hyperbolic modes occupy the phase space where $\operatorname{Re}\left[\text{K}D\right]\in\left(0,\,\pi\right)$,
and have to fulfill the condition in Eq. \eqref{eq:3.1} in order to
be stable. Combining these, the two boundaries of the modes are:

\begin{equation}
\begin{cases}
\operatorname{Re}\left[\text{K}D\right]=\operatorname{Im}\left[\text{K}D\right]=0 & \left(a\right)\\
\operatorname{Re}\left[\text{K}D\right]=\operatorname{Im}\left[\text{K}D\right]=\pi & \left(b\right)
\end{cases}
\end{equation}

Thus, using these rules, the conditions for $\sigma_{s}^{''}(q,\omega)$
can be derived from the Eq. \eqref{eq:B.2}:

\begin{equation}
\begin{cases}
\sigma_{s}^{''}(q,\omega)=\frac{2\varepsilon_{1}}{Z_{0}a}\tanh\left(\frac{k_{0}d_{1}a}{2}\right) & \left(a\right)\\
\sigma_{s}^{''}(q,\omega)=\frac{2\varepsilon_{1}}{Z_{0}a}\frac{\cosh\left(k_{0}d_{1}a\right)+\cosh\left(\pi\right)}{\sinh\left(k_{0}d_{1}a\right)} & \left(b\right)
\end{cases}\label{eq:B.3}
\end{equation}
Finding the values of $(q,\omega)$, at which the above equalities
are satisfied, will define the region $\varepsilon_{x-eff}<0$.

The same procedure can be used to find the boundary of undamped hyperbolic
modes, where $\operatorname{Re}\left[\text{K}D\right]=\pi$, and $\operatorname{Im}\left[\text{K}D\right]=0$:

\begin{equation}
\sigma_{s}^{''}(q,\omega)=\frac{2\varepsilon_{1}}{Z_{0}a}\left(\coth\left(\frac{k_{0}d_{1}a}{2}\right)+\frac{1}{\sinh\left(\frac{k_{0}d_{1}a}{2}\right)}\right)\label{eq:B.4}
\end{equation}

The conditions $(a)$ and $(b)$ of the Eqs. \eqref{eq:B.3}, bounding
the region of $\varepsilon_{x-eff}<0$, are plotted as black solid
and black dashed lines respectively in the Figs. \ref{fig:epsilon-x-eff a}
and \ref{fig:epsilon-x-eff c}. Within that region, the
undamped modes have their $k_{x}/k_{0}$ values bound by the dotted
curve, given by the Eq. \eqref{eq:B.4}. All three curves tend to the
single graphene sheet dispersion curve, shown as a green line. As
the distance between the sheets increases ($d_{1}\rightarrow\infty$),
the region of $\varepsilon_{x-eff}<0$ becomes smaller, converging
to $\sigma_{s}^{''}(q,\omega)=\frac{2\varepsilon_{1}}{Z_{0}a}$. Expectedly,
this coincides with the single-sheet plasmon dispersion.


\begin{thebibliography}{43}%
\makeatletter
\providecommand \@ifxundefined [1]{%
 \@ifx{#1\undefined}
}%
\providecommand \@ifnum [1]{%
 \ifnum #1\expandafter \@firstoftwo
 \else \expandafter \@secondoftwo
 \fi
}%
\providecommand \@ifx [1]{%
 \ifx #1\expandafter \@firstoftwo
 \else \expandafter \@secondoftwo
 \fi
}%
\providecommand \natexlab [1]{#1}%
\providecommand \enquote  [1]{``#1''}%
\providecommand \bibnamefont  [1]{#1}%
\providecommand \bibfnamefont [1]{#1}%
\providecommand \citenamefont [1]{#1}%
\providecommand \href@noop [0]{\@secondoftwo}%
\providecommand \href [0]{\begingroup \@sanitize@url \@href}%
\providecommand \@href[1]{\@@startlink{#1}\@@href}%
\providecommand \@@href[1]{\endgroup#1\@@endlink}%
\providecommand \@sanitize@url [0]{\catcode `\\12\catcode `\$12\catcode
  `\&12\catcode `\#12\catcode `\^12\catcode `\_12\catcode `\%12\relax}%
\providecommand \@@startlink[1]{}%
\providecommand \@@endlink[0]{}%
\providecommand \url  [0]{\begingroup\@sanitize@url \@url }%
\providecommand \@url [1]{\endgroup\@href {#1}{\urlprefix }}%
\providecommand \urlprefix  [0]{URL }%
\providecommand \Eprint [0]{\href }%
\providecommand \doibase [0]{http://dx.doi.org/}%
\providecommand \selectlanguage [0]{\@gobble}%
\providecommand \bibinfo  [0]{\@secondoftwo}%
\providecommand \bibfield  [0]{\@secondoftwo}%
\providecommand \translation [1]{[#1]}%
\providecommand \BibitemOpen [0]{}%
\providecommand \bibitemStop [0]{}%
\providecommand \bibitemNoStop [0]{.\EOS\space}%
\providecommand \EOS [0]{\spacefactor3000\relax}%
\providecommand \BibitemShut  [1]{\csname bibitem#1\endcsname}%
\let\auto@bib@innerbib\@empty
%</preamble>
\bibitem [{\citenamefont {Shung}(1986)}]{Shung1986}%
  \BibitemOpen
  \bibfield  {author} {\bibinfo {author} {\bibfnamefont {K.~W.~K.}\
  \bibnamefont {Shung}},\ }\href {\doibase 10.1103/PhysRevB.34.979} {\bibfield
  {journal} {\bibinfo  {journal} {Phys. Rev. B}\ }\textbf {\bibinfo {volume}
  {34}},\ \bibinfo {pages} {979} (\bibinfo {year} {1986})}\BibitemShut
  {NoStop}%
\bibitem [{\citenamefont {Yan}\ \emph {et~al.}(2012)\citenamefont {Yan},
  \citenamefont {Li}, \citenamefont {Chandra}, \citenamefont {Tulevski},
  \citenamefont {Wu}, \citenamefont {Freitag}, \citenamefont {Zhu},
  \citenamefont {Avouris},\ and\ \citenamefont {Xia}}]{Yan2012}%
  \BibitemOpen
  \bibfield  {author} {\bibinfo {author} {\bibfnamefont {H.}~\bibnamefont
  {Yan}}, \bibinfo {author} {\bibfnamefont {X.}~\bibnamefont {Li}}, \bibinfo
  {author} {\bibfnamefont {B.}~\bibnamefont {Chandra}}, \bibinfo {author}
  {\bibfnamefont {G.}~\bibnamefont {Tulevski}}, \bibinfo {author}
  {\bibfnamefont {Y.}~\bibnamefont {Wu}}, \bibinfo {author} {\bibfnamefont
  {M.}~\bibnamefont {Freitag}}, \bibinfo {author} {\bibfnamefont
  {W.}~\bibnamefont {Zhu}}, \bibinfo {author} {\bibfnamefont {P.}~\bibnamefont
  {Avouris}}, \ and\ \bibinfo {author} {\bibfnamefont {F.}~\bibnamefont
  {Xia}},\ }\href {\doibase 10.1038/nnano.2012.59} {\bibfield  {journal}
  {\bibinfo  {journal} {Nat. Nanotechnol.}\ }\textbf {\bibinfo {volume} {7}},\
  \bibinfo {pages} {330} (\bibinfo {year} {2012})}\BibitemShut {NoStop}%
\bibitem [{\citenamefont {Iorsh}\ \emph {et~al.}(2013)\citenamefont {Iorsh},
  \citenamefont {Mukhin}, \citenamefont {Shadrivov}, \citenamefont {Belov},\
  and\ \citenamefont {Kivshar}}]{Iorsh2013}%
  \BibitemOpen
  \bibfield  {author} {\bibinfo {author} {\bibfnamefont {I.~V.}\ \bibnamefont
  {Iorsh}}, \bibinfo {author} {\bibfnamefont {I.~S.}\ \bibnamefont {Mukhin}},
  \bibinfo {author} {\bibfnamefont {I.~V.}\ \bibnamefont {Shadrivov}}, \bibinfo
  {author} {\bibfnamefont {P.~A.}\ \bibnamefont {Belov}}, \ and\ \bibinfo
  {author} {\bibfnamefont {Y.~S.}\ \bibnamefont {Kivshar}},\ }\href {\doibase
  10.1103/PhysRevB.87.075416} {\bibfield  {journal} {\bibinfo  {journal} {Phys.
  Rev. B}\ }\textbf {\bibinfo {volume} {87}},\ \bibinfo {pages} {075416}
  (\bibinfo {year} {2013})}\BibitemShut {NoStop}%
\bibitem [{\citenamefont {Sreekanth}\ \emph {et~al.}(2013)\citenamefont
  {Sreekanth}, \citenamefont {{De Luca}},\ and\ \citenamefont
  {Strangi}}]{Sreekanth2013}%
  \BibitemOpen
  \bibfield  {author} {\bibinfo {author} {\bibfnamefont {K.~V.}\ \bibnamefont
  {Sreekanth}}, \bibinfo {author} {\bibfnamefont {A.}~\bibnamefont {{De
  Luca}}}, \ and\ \bibinfo {author} {\bibfnamefont {G.}~\bibnamefont
  {Strangi}},\ }\href {\doibase 10.1063/1.4813477} {\bibfield  {journal}
  {\bibinfo  {journal} {Appl. Phys. Lett.}\ }\textbf {\bibinfo {volume}
  {103}},\ \bibinfo {pages} {023107} (\bibinfo {year} {2013})}\BibitemShut
  {NoStop}%
\bibitem [{\citenamefont {Othman}\ \emph {et~al.}(2013)\citenamefont {Othman},
  \citenamefont {Guclu},\ and\ \citenamefont
  {Capolino}}]{MohamedA.K.Othman2013}%
  \BibitemOpen
  \bibfield  {author} {\bibinfo {author} {\bibfnamefont {M.~A.~K.}\
  \bibnamefont {Othman}}, \bibinfo {author} {\bibfnamefont {C.}~\bibnamefont
  {Guclu}}, \ and\ \bibinfo {author} {\bibfnamefont {F.}~\bibnamefont
  {Capolino}},\ }\href {\doibase 10.1364/OE.21.007614} {\bibfield  {journal}
  {\bibinfo  {journal} {Opt. Express}\ }\textbf {\bibinfo {volume} {21}},\
  \bibinfo {pages} {7614} (\bibinfo {year} {2013})}\BibitemShut {NoStop}%
\bibitem [{\citenamefont {Poddubny}\ \emph {et~al.}(2013)\citenamefont
  {Poddubny}, \citenamefont {Iorsh}, \citenamefont {Belov},\ and\ \citenamefont
  {Kivshar}}]{Poddubny2013}%
  \BibitemOpen
  \bibfield  {author} {\bibinfo {author} {\bibfnamefont {A.}~\bibnamefont
  {Poddubny}}, \bibinfo {author} {\bibfnamefont {I.}~\bibnamefont {Iorsh}},
  \bibinfo {author} {\bibfnamefont {P.}~\bibnamefont {Belov}}, \ and\ \bibinfo
  {author} {\bibfnamefont {Y.}~\bibnamefont {Kivshar}},\ }\href {\doibase
  10.1038/nphoton.2013.243} {\bibfield  {journal} {\bibinfo  {journal} {Nat.
  Photonics}\ }\textbf {\bibinfo {volume} {7}},\ \bibinfo {pages} {948}
  (\bibinfo {year} {2013})}\BibitemShut {NoStop}%
\bibitem [{\citenamefont {Ferrari}\ \emph {et~al.}(2015)\citenamefont
  {Ferrari}, \citenamefont {Wu}, \citenamefont {Lepage}, \citenamefont
  {Zhang},\ and\ \citenamefont {Liu}}]{Ferrari2015}%
  \BibitemOpen
  \bibfield  {author} {\bibinfo {author} {\bibfnamefont {L.}~\bibnamefont
  {Ferrari}}, \bibinfo {author} {\bibfnamefont {C.}~\bibnamefont {Wu}},
  \bibinfo {author} {\bibfnamefont {D.}~\bibnamefont {Lepage}}, \bibinfo
  {author} {\bibfnamefont {X.}~\bibnamefont {Zhang}}, \ and\ \bibinfo {author}
  {\bibfnamefont {Z.}~\bibnamefont {Liu}},\ }\href {\doibase
  10.1016/j.pquantelec.2014.10.001} {\bibfield  {journal} {\bibinfo  {journal}
  {Prog. Quantum Electron.}\ }\textbf {\bibinfo {volume} {40}},\ \bibinfo
  {pages} {1} (\bibinfo {year} {2015})}\BibitemShut {NoStop}%
\bibitem [{\citenamefont {Chang}\ \emph {et~al.}(2016)\citenamefont {Chang},
  \citenamefont {Liu}, \citenamefont {Liu}, \citenamefont {Zhang},
  \citenamefont {Marder}, \citenamefont {Narimanov}, \citenamefont {Zhong},\
  and\ \citenamefont {Norris}}]{Chang2016}%
  \BibitemOpen
  \bibfield  {author} {\bibinfo {author} {\bibfnamefont {Y.-C.}\ \bibnamefont
  {Chang}}, \bibinfo {author} {\bibfnamefont {C.-H.}\ \bibnamefont {Liu}},
  \bibinfo {author} {\bibfnamefont {C.-H.}\ \bibnamefont {Liu}}, \bibinfo
  {author} {\bibfnamefont {S.}~\bibnamefont {Zhang}}, \bibinfo {author}
  {\bibfnamefont {S.~R.}\ \bibnamefont {Marder}}, \bibinfo {author}
  {\bibfnamefont {E.~E.}\ \bibnamefont {Narimanov}}, \bibinfo {author}
  {\bibfnamefont {Z.}~\bibnamefont {Zhong}}, \ and\ \bibinfo {author}
  {\bibfnamefont {T.~B.}\ \bibnamefont {Norris}},\ }\href {\doibase
  10.1038/ncomms10568} {\bibfield  {journal} {\bibinfo  {journal} {Nat.
  Commun.}\ }\textbf {\bibinfo {volume} {7}},\ \bibinfo {pages} {10568}
  (\bibinfo {year} {2016})}\BibitemShut {NoStop}%
\bibitem [{\citenamefont {Novoselov}\ \emph {et~al.}(2016)\citenamefont
  {Novoselov}, \citenamefont {Mishchenko}, \citenamefont {Carvalho},\ and\
  \citenamefont {{Castro Neto}}}]{Novoselov2016}%
  \BibitemOpen
  \bibfield  {author} {\bibinfo {author} {\bibfnamefont {K.~S.}\ \bibnamefont
  {Novoselov}}, \bibinfo {author} {\bibfnamefont {A.}~\bibnamefont
  {Mishchenko}}, \bibinfo {author} {\bibfnamefont {A.}~\bibnamefont
  {Carvalho}}, \ and\ \bibinfo {author} {\bibfnamefont {A.~H.}\ \bibnamefont
  {{Castro Neto}}},\ }\href {\doibase 10.1126/science.aac9439} {\bibfield
  {journal} {\bibinfo  {journal} {Science}\ }\textbf {\bibinfo {volume} {353}}
  (\bibinfo {year} {2016}),\ 10.1126/science.aac9439}\BibitemShut {NoStop}%
\bibitem [{\citenamefont {Vorobev}\ and\ \citenamefont
  {Tyukhtin}(2012)}]{Veselago1968}%
  \BibitemOpen
  \bibfield  {author} {\bibinfo {author} {\bibfnamefont {V.~V.}\ \bibnamefont
  {Vorobev}}\ and\ \bibinfo {author} {\bibfnamefont {A.~V.}\ \bibnamefont
  {Tyukhtin}},\ }\href {\doibase 10.1103/PhysRevLett.108.184801} {\bibfield
  {journal} {\bibinfo  {journal} {Phys. Rev. Lett.}\ }\textbf {\bibinfo
  {volume} {108}},\ \bibinfo {pages} {184801} (\bibinfo {year}
  {2012})}\BibitemShut {NoStop}%
\bibitem [{\citenamefont {Pendry}(2000)}]{Pendry2000}%
  \BibitemOpen
  \bibfield  {author} {\bibinfo {author} {\bibfnamefont {J.~B.}\ \bibnamefont
  {Pendry}},\ }\href {\doibase 10.1103/PhysRevLett.85.3966} {\bibfield
  {journal} {\bibinfo  {journal} {Phys. Rev. Lett.}\ }\textbf {\bibinfo
  {volume} {85}},\ \bibinfo {pages} {3966} (\bibinfo {year}
  {2000})}\BibitemShut {NoStop}%
\bibitem [{\citenamefont {Belov}\ and\ \citenamefont {Hao}(2006)}]{Belov2006}%
  \BibitemOpen
  \bibfield  {author} {\bibinfo {author} {\bibfnamefont {P.~a.}\ \bibnamefont
  {Belov}}\ and\ \bibinfo {author} {\bibfnamefont {Y.}~\bibnamefont {Hao}},\
  }\href {\doibase 10.1103/PhysRevB.73.113110} {\bibfield  {journal} {\bibinfo
  {journal} {Phys. Rev. B}\ }\textbf {\bibinfo {volume} {73}},\ \bibinfo
  {pages} {113110} (\bibinfo {year} {2006})}\BibitemShut {NoStop}%
\bibitem [{\citenamefont {Salandrino}\ and\ \citenamefont
  {Engheta}(2006)}]{Salandrino2006}%
  \BibitemOpen
  \bibfield  {author} {\bibinfo {author} {\bibfnamefont {A.}~\bibnamefont
  {Salandrino}}\ and\ \bibinfo {author} {\bibfnamefont {N.}~\bibnamefont
  {Engheta}},\ }\href {\doibase 10.1103/PhysRevB.74.075103} {\bibfield
  {journal} {\bibinfo  {journal} {Phys. Rev. B}\ }\textbf {\bibinfo {volume}
  {74}},\ \bibinfo {pages} {075103} (\bibinfo {year} {2006})}\BibitemShut
  {NoStop}%
\bibitem [{\citenamefont {Wood}\ \emph {et~al.}(2006)\citenamefont {Wood},
  \citenamefont {Pendry},\ and\ \citenamefont {Tsai}}]{Wood2006}%
  \BibitemOpen
  \bibfield  {author} {\bibinfo {author} {\bibfnamefont {B.}~\bibnamefont
  {Wood}}, \bibinfo {author} {\bibfnamefont {J.~B.}\ \bibnamefont {Pendry}}, \
  and\ \bibinfo {author} {\bibfnamefont {D.~P.}\ \bibnamefont {Tsai}},\ }\href
  {\doibase 10.1103/PhysRevB.74.115116} {\bibfield  {journal} {\bibinfo
  {journal} {Phys. Rev. B}\ }\textbf {\bibinfo {volume} {74}},\ \bibinfo
  {pages} {115116} (\bibinfo {year} {2006})}\BibitemShut {NoStop}%
\bibitem [{\citenamefont {Rho}\ \emph {et~al.}(2010)\citenamefont {Rho},
  \citenamefont {Ye}, \citenamefont {Xiong}, \citenamefont {Yin}, \citenamefont
  {Liu}, \citenamefont {Choi}, \citenamefont {Bartal},\ and\ \citenamefont
  {Zhang}}]{Rho2010}%
  \BibitemOpen
  \bibfield  {author} {\bibinfo {author} {\bibfnamefont {J.}~\bibnamefont
  {Rho}}, \bibinfo {author} {\bibfnamefont {Z.}~\bibnamefont {Ye}}, \bibinfo
  {author} {\bibfnamefont {Y.}~\bibnamefont {Xiong}}, \bibinfo {author}
  {\bibfnamefont {X.}~\bibnamefont {Yin}}, \bibinfo {author} {\bibfnamefont
  {Z.}~\bibnamefont {Liu}}, \bibinfo {author} {\bibfnamefont {H.}~\bibnamefont
  {Choi}}, \bibinfo {author} {\bibfnamefont {G.}~\bibnamefont {Bartal}}, \ and\
  \bibinfo {author} {\bibfnamefont {X.}~\bibnamefont {Zhang}},\ }\href
  {\doibase 10.1038/ncomms1148} {\bibfield  {journal} {\bibinfo  {journal}
  {Nat. Commun.}\ }\textbf {\bibinfo {volume} {1}},\ \bibinfo {pages} {143}
  (\bibinfo {year} {2010})}\BibitemShut {NoStop}%
\bibitem [{\citenamefont {Andryieuski}\ \emph {et~al.}(2012)\citenamefont
  {Andryieuski}, \citenamefont {Lavrinenko},\ and\ \citenamefont
  {Chigrin}}]{Andryieuski2012}%
  \BibitemOpen
  \bibfield  {author} {\bibinfo {author} {\bibfnamefont {A.}~\bibnamefont
  {Andryieuski}}, \bibinfo {author} {\bibfnamefont {A.~V.}\ \bibnamefont
  {Lavrinenko}}, \ and\ \bibinfo {author} {\bibfnamefont {D.~N.}\ \bibnamefont
  {Chigrin}},\ }\href {\doibase 10.1103/PhysRevB.86.121108} {\bibfield
  {journal} {\bibinfo  {journal} {Phys. Rev. B}\ }\textbf {\bibinfo {volume}
  {86}},\ \bibinfo {pages} {121108} (\bibinfo {year} {2012})}\BibitemShut
  {NoStop}%
\bibitem [{\citenamefont {Yang}\ \emph {et~al.}(2010)\citenamefont {Yang},
  \citenamefont {Yin}, \citenamefont {Xu}, \citenamefont {Feng},\ and\
  \citenamefont {Du}}]{Galfsky2015}%
  \BibitemOpen
  \bibfield  {author} {\bibinfo {author} {\bibfnamefont {W.~L.}\ \bibnamefont
  {Yang}}, \bibinfo {author} {\bibfnamefont {Z.~Q.}\ \bibnamefont {Yin}},
  \bibinfo {author} {\bibfnamefont {Z.~Y.}\ \bibnamefont {Xu}}, \bibinfo
  {author} {\bibfnamefont {M.}~\bibnamefont {Feng}}, \ and\ \bibinfo {author}
  {\bibfnamefont {J.~F.}\ \bibnamefont {Du}},\ }\href {\doibase
  10.1063/1.3455891} {\bibfield  {journal} {\bibinfo  {journal} {Appl. Phys.
  Lett.}\ }\textbf {\bibinfo {volume} {96}},\ \bibinfo {pages} {241113}
  (\bibinfo {year} {2010})}\BibitemShut {NoStop}%
\bibitem [{\citenamefont {Tumkur}\ \emph {et~al.}(2011)\citenamefont {Tumkur},
  \citenamefont {Zhu}, \citenamefont {Black}, \citenamefont {Barnakov},
  \citenamefont {Bonner},\ and\ \citenamefont {Noginov}}]{Tumkur2011}%
  \BibitemOpen
  \bibfield  {author} {\bibinfo {author} {\bibfnamefont {T.}~\bibnamefont
  {Tumkur}}, \bibinfo {author} {\bibfnamefont {G.}~\bibnamefont {Zhu}},
  \bibinfo {author} {\bibfnamefont {P.}~\bibnamefont {Black}}, \bibinfo
  {author} {\bibfnamefont {Y.~a.}\ \bibnamefont {Barnakov}}, \bibinfo {author}
  {\bibfnamefont {C.~E.}\ \bibnamefont {Bonner}}, \ and\ \bibinfo {author}
  {\bibfnamefont {M.~a.}\ \bibnamefont {Noginov}},\ }\href {\doibase
  10.1063/1.3631723} {\bibfield  {journal} {\bibinfo  {journal} {Appl. Phys.
  Lett.}\ }\textbf {\bibinfo {volume} {99}},\ \bibinfo {pages} {151115}
  (\bibinfo {year} {2011})}\BibitemShut {NoStop}%
\bibitem [{\citenamefont {Xiong}\ \emph {et~al.}(2009)\citenamefont {Xiong},
  \citenamefont {Liu},\ and\ \citenamefont {Zhang}}]{Xiong2009}%
  \BibitemOpen
  \bibfield  {author} {\bibinfo {author} {\bibfnamefont {Y.}~\bibnamefont
  {Xiong}}, \bibinfo {author} {\bibfnamefont {Z.}~\bibnamefont {Liu}}, \ and\
  \bibinfo {author} {\bibfnamefont {X.}~\bibnamefont {Zhang}},\ }\href
  {\doibase 10.1063/1.3141457} {\bibfield  {journal} {\bibinfo  {journal}
  {Appl. Phys. Lett.}\ }\textbf {\bibinfo {volume} {94}},\ \bibinfo {pages}
  {203108} (\bibinfo {year} {2009})}\BibitemShut {NoStop}%
\bibitem [{\citenamefont {Smolyaninov}\ and\ \citenamefont
  {Hung}(2011)}]{Smolyaninov2011}%
  \BibitemOpen
  \bibfield  {author} {\bibinfo {author} {\bibfnamefont {I.~I.}\ \bibnamefont
  {Smolyaninov}}\ and\ \bibinfo {author} {\bibfnamefont {Y.-j.}\ \bibnamefont
  {Hung}},\ }\href {\doibase 10.1364/JOSAB.28.001591} {\bibfield  {journal}
  {\bibinfo  {journal} {Journal of the Optical Society of America B}\ }\textbf
  {\bibinfo {volume} {28}},\ \bibinfo {pages} {1591} (\bibinfo {year}
  {2011})}\BibitemShut {NoStop}%
\bibitem [{\citenamefont {Dai}\ \emph {et~al.}(2015)\citenamefont {Dai},
  \citenamefont {Ma}, \citenamefont {Liu}, \citenamefont {Andersen},
  \citenamefont {Fei}, \citenamefont {Goldflam}, \citenamefont {Wagner},
  \citenamefont {Watanabe}, \citenamefont {Taniguchi}, \citenamefont
  {Thiemens}, \citenamefont {Keilmann}, \citenamefont {Janssen}, \citenamefont
  {Zhu}, \citenamefont {Jarillo-Herrero}, \citenamefont {Fogler},\ and\
  \citenamefont {Basov}}]{Dai2015a}%
  \BibitemOpen
  \bibfield  {author} {\bibinfo {author} {\bibfnamefont {S.}~\bibnamefont
  {Dai}}, \bibinfo {author} {\bibfnamefont {Q.}~\bibnamefont {Ma}}, \bibinfo
  {author} {\bibfnamefont {M.~K.}\ \bibnamefont {Liu}}, \bibinfo {author}
  {\bibfnamefont {T.}~\bibnamefont {Andersen}}, \bibinfo {author}
  {\bibfnamefont {Z.}~\bibnamefont {Fei}}, \bibinfo {author} {\bibfnamefont
  {M.~D.}\ \bibnamefont {Goldflam}}, \bibinfo {author} {\bibfnamefont
  {M.}~\bibnamefont {Wagner}}, \bibinfo {author} {\bibfnamefont
  {K.}~\bibnamefont {Watanabe}}, \bibinfo {author} {\bibfnamefont
  {T.}~\bibnamefont {Taniguchi}}, \bibinfo {author} {\bibfnamefont
  {M.}~\bibnamefont {Thiemens}}, \bibinfo {author} {\bibfnamefont
  {F.}~\bibnamefont {Keilmann}}, \bibinfo {author} {\bibfnamefont {G.~C.
  a.~M.}\ \bibnamefont {Janssen}}, \bibinfo {author} {\bibfnamefont {S.-E.}\
  \bibnamefont {Zhu}}, \bibinfo {author} {\bibfnamefont {P.}~\bibnamefont
  {Jarillo-Herrero}}, \bibinfo {author} {\bibfnamefont {M.~M.}\ \bibnamefont
  {Fogler}}, \ and\ \bibinfo {author} {\bibfnamefont {D.~N.}\ \bibnamefont
  {Basov}},\ }\href {\doibase 10.1038/nnano.2015.131} {\bibfield  {journal}
  {\bibinfo  {journal} {Nat. Nanotechnol.}\ }\textbf {\bibinfo {volume} {10}},\
  \bibinfo {pages} {682} (\bibinfo {year} {2015})}\BibitemShut {NoStop}%
\bibitem [{\citenamefont {Orlov}\ \emph {et~al.}(2011)\citenamefont {Orlov},
  \citenamefont {Voroshilov}, \citenamefont {Belov},\ and\ \citenamefont
  {Kivshar}}]{Orlov2011}%
  \BibitemOpen
  \bibfield  {author} {\bibinfo {author} {\bibfnamefont {A.~A.}\ \bibnamefont
  {Orlov}}, \bibinfo {author} {\bibfnamefont {P.~M.}\ \bibnamefont
  {Voroshilov}}, \bibinfo {author} {\bibfnamefont {P.~A.}\ \bibnamefont
  {Belov}}, \ and\ \bibinfo {author} {\bibfnamefont {Y.~S.}\ \bibnamefont
  {Kivshar}},\ }\href {\doibase 10.1103/PhysRevB.84.045424} {\bibfield
  {journal} {\bibinfo  {journal} {Phys. Rev. B}\ }\textbf {\bibinfo {volume}
  {84}},\ \bibinfo {pages} {045424} (\bibinfo {year} {2011})}\BibitemShut
  {NoStop}%
\bibitem [{\citenamefont {Orlov}\ \emph {et~al.}(2014)\citenamefont {Orlov},
  \citenamefont {Zhukovsky}, \citenamefont {Iorsh},\ and\ \citenamefont
  {Belov}}]{Orlov2014}%
  \BibitemOpen
  \bibfield  {author} {\bibinfo {author} {\bibfnamefont {A.~A.}\ \bibnamefont
  {Orlov}}, \bibinfo {author} {\bibfnamefont {S.~V.}\ \bibnamefont
  {Zhukovsky}}, \bibinfo {author} {\bibfnamefont {I.~V.}\ \bibnamefont
  {Iorsh}}, \ and\ \bibinfo {author} {\bibfnamefont {P.~A.}\ \bibnamefont
  {Belov}},\ }\href {\doibase 10.1016/j.photonics.2014.03.003} {\bibfield
  {journal} {\bibinfo  {journal} {Photonics Nanostruct. Fundam. Appl.}\
  }\textbf {\bibinfo {volume} {12}},\ \bibinfo {pages} {213} (\bibinfo {year}
  {2014})}\BibitemShut {NoStop}%
\bibitem [{\citenamefont {Page}\ \emph {et~al.}(2015)\citenamefont {Page},
  \citenamefont {Ballout}, \citenamefont {Hess},\ and\ \citenamefont
  {Hamm}}]{Page2015}%
  \BibitemOpen
  \bibfield  {author} {\bibinfo {author} {\bibfnamefont {A.~F.}\ \bibnamefont
  {Page}}, \bibinfo {author} {\bibfnamefont {F.}~\bibnamefont {Ballout}},
  \bibinfo {author} {\bibfnamefont {O.}~\bibnamefont {Hess}}, \ and\ \bibinfo
  {author} {\bibfnamefont {J.~M.}\ \bibnamefont {Hamm}},\ }\href {\doibase
  10.1103/PhysRevB.91.075404} {\bibfield  {journal} {\bibinfo  {journal} {Phys.
  Rev. B}\ }\textbf {\bibinfo {volume} {91}},\ \bibinfo {pages} {075404}
  (\bibinfo {year} {2015})}\BibitemShut {NoStop}%
\bibitem [{\citenamefont {Savelev}\ \emph {et~al.}(2013)\citenamefont
  {Savelev}, \citenamefont {Shadrivov}, \citenamefont {Belov}, \citenamefont
  {Rosanov}, \citenamefont {Fedorov}, \citenamefont {Sukhorukov},\ and\
  \citenamefont {Kivshar}}]{Savelev2013}%
  \BibitemOpen
  \bibfield  {author} {\bibinfo {author} {\bibfnamefont {R.~S.}\ \bibnamefont
  {Savelev}}, \bibinfo {author} {\bibfnamefont {I.~V.}\ \bibnamefont
  {Shadrivov}}, \bibinfo {author} {\bibfnamefont {P.~a.}\ \bibnamefont
  {Belov}}, \bibinfo {author} {\bibfnamefont {N.~N.}\ \bibnamefont {Rosanov}},
  \bibinfo {author} {\bibfnamefont {S.~V.}\ \bibnamefont {Fedorov}}, \bibinfo
  {author} {\bibfnamefont {A.~a.}\ \bibnamefont {Sukhorukov}}, \ and\ \bibinfo
  {author} {\bibfnamefont {Y.~S.}\ \bibnamefont {Kivshar}},\ }\href {\doibase
  10.1103/PhysRevB.87.115139} {\bibfield  {journal} {\bibinfo  {journal} {Phys.
  Rev. B}\ }\textbf {\bibinfo {volume} {87}},\ \bibinfo {pages} {115139}
  (\bibinfo {year} {2013})}\BibitemShut {NoStop}%
\bibitem [{\citenamefont {Pyatkovskiy}(2009)}]{Pyatkovskiy2009a}%
  \BibitemOpen
  \bibfield  {author} {\bibinfo {author} {\bibfnamefont {P.~K.}\ \bibnamefont
  {Pyatkovskiy}},\ }\href {\doibase 10.1088/0953-8984/21/2/025506} {\bibfield
  {journal} {\bibinfo  {journal} {J. Phys.: Condens. Matter}\ }\textbf
  {\bibinfo {volume} {21}},\ \bibinfo {pages} {025506} (\bibinfo {year}
  {2009})}\BibitemShut {NoStop}%
\bibitem [{\citenamefont {Wunsch}\ \emph {et~al.}(2006)\citenamefont {Wunsch},
  \citenamefont {Stauber}, \citenamefont {Sols},\ and\ \citenamefont
  {Guinea}}]{Wunsch2006}%
  \BibitemOpen
  \bibfield  {author} {\bibinfo {author} {\bibfnamefont {B.}~\bibnamefont
  {Wunsch}}, \bibinfo {author} {\bibfnamefont {T.}~\bibnamefont {Stauber}},
  \bibinfo {author} {\bibfnamefont {F.}~\bibnamefont {Sols}}, \ and\ \bibinfo
  {author} {\bibfnamefont {F.}~\bibnamefont {Guinea}},\ }\href {\doibase
  10.1088/1367-2630/8/12/318} {\bibfield  {journal} {\bibinfo  {journal} {New
  J. Phys.}\ }\textbf {\bibinfo {volume} {8}},\ \bibinfo {pages} {318}
  (\bibinfo {year} {2006})}\BibitemShut {NoStop}%
\bibitem [{\citenamefont {Hwang}\ and\ \citenamefont {{Das
  Sarma}}(2007)}]{Hwang2007}%
  \BibitemOpen
  \bibfield  {author} {\bibinfo {author} {\bibfnamefont {E.~H.}\ \bibnamefont
  {Hwang}}\ and\ \bibinfo {author} {\bibfnamefont {S.}~\bibnamefont {{Das
  Sarma}}},\ }\href {\doibase 10.1103/PhysRevB.75.205418} {\bibfield  {journal}
  {\bibinfo  {journal} {Phys. Rev. B}\ }\textbf {\bibinfo {volume} {75}},\
  \bibinfo {pages} {205418} (\bibinfo {year} {2007})}\BibitemShut {NoStop}%
\bibitem [{\citenamefont {Watanabe}\ \emph {et~al.}(2013)\citenamefont
  {Watanabe}, \citenamefont {Fukushima}, \citenamefont {Yabe}, \citenamefont
  {{Boubanga Tombet}}, \citenamefont {Satou}, \citenamefont {Dubinov},
  \citenamefont {Aleshkin}, \citenamefont {Mitin}, \citenamefont {Ryzhii},\
  and\ \citenamefont {Otsuji}}]{Watanabe2013}%
  \BibitemOpen
  \bibfield  {author} {\bibinfo {author} {\bibfnamefont {T.}~\bibnamefont
  {Watanabe}}, \bibinfo {author} {\bibfnamefont {T.}~\bibnamefont {Fukushima}},
  \bibinfo {author} {\bibfnamefont {Y.}~\bibnamefont {Yabe}}, \bibinfo {author}
  {\bibfnamefont {S.~A.}\ \bibnamefont {{Boubanga Tombet}}}, \bibinfo {author}
  {\bibfnamefont {A.}~\bibnamefont {Satou}}, \bibinfo {author} {\bibfnamefont
  {A.~A.}\ \bibnamefont {Dubinov}}, \bibinfo {author} {\bibfnamefont {V.~Y.}\
  \bibnamefont {Aleshkin}}, \bibinfo {author} {\bibfnamefont {V.}~\bibnamefont
  {Mitin}}, \bibinfo {author} {\bibfnamefont {V.}~\bibnamefont {Ryzhii}}, \
  and\ \bibinfo {author} {\bibfnamefont {T.}~\bibnamefont {Otsuji}},\ }\href
  {\doibase 10.1088/1367-2630/15/7/075003} {\bibfield  {journal} {\bibinfo
  {journal} {New J. Phys.}\ }\textbf {\bibinfo {volume} {15}},\ \bibinfo
  {pages} {075003} (\bibinfo {year} {2013})}\BibitemShut {NoStop}%
\bibitem [{\citenamefont {Dubinov}\ \emph {et~al.}(2011)\citenamefont
  {Dubinov}, \citenamefont {Aleshkin}, \citenamefont {Mitin}, \citenamefont
  {Otsuji},\ and\ \citenamefont {Ryzhii}}]{Dubinov2011}%
  \BibitemOpen
  \bibfield  {author} {\bibinfo {author} {\bibfnamefont {A.~A.}\ \bibnamefont
  {Dubinov}}, \bibinfo {author} {\bibfnamefont {V.~Y.}\ \bibnamefont
  {Aleshkin}}, \bibinfo {author} {\bibfnamefont {V.}~\bibnamefont {Mitin}},
  \bibinfo {author} {\bibfnamefont {T.}~\bibnamefont {Otsuji}}, \ and\ \bibinfo
  {author} {\bibfnamefont {V.}~\bibnamefont {Ryzhii}},\ }\href {\doibase
  10.1088/0953-8984/23/14/145302} {\bibfield  {journal} {\bibinfo  {journal}
  {J. Phys.: Condens. Matter}\ }\textbf {\bibinfo {volume} {23}},\ \bibinfo
  {pages} {145302} (\bibinfo {year} {2011})}\BibitemShut {NoStop}%
\bibitem [{\citenamefont {Stern}(1967)}]{Stern1967}%
  \BibitemOpen
  \bibfield  {author} {\bibinfo {author} {\bibfnamefont {F.}~\bibnamefont
  {Stern}},\ }\href {\doibase 10.1103/PhysRevLett.18.546} {\bibfield  {journal}
  {\bibinfo  {journal} {Phys. Rev. Lett.}\ }\textbf {\bibinfo {volume} {18}},\
  \bibinfo {pages} {546} (\bibinfo {year} {1967})}\BibitemShut {NoStop}%
\bibitem [{\citenamefont {Page}(2016)}]{Page2016}%
  \BibitemOpen
  \bibfield  {author} {\bibinfo {author} {\bibfnamefont {A.~F.}\ \bibnamefont
  {Page}},\ }\emph {\bibinfo {title} {{Surface plasmon emission and dynamics in
  active planar media}}},\ \href
  {spiral.imperial.ac.uk:8443/handle/10044/1/30760} {Ph.D. thesis},\ \bibinfo
  {school} {Imperial College London} (\bibinfo {year} {2016})\BibitemShut
  {NoStop}%
\bibitem [{\citenamefont {Page}\ \emph {et~al.}(2018)\citenamefont {Page},
  \citenamefont {Hamm},\ and\ \citenamefont {Hess}}]{Page2018}%
  \BibitemOpen
  \bibfield  {author} {\bibinfo {author} {\bibfnamefont {A.~F.}\ \bibnamefont
  {Page}}, \bibinfo {author} {\bibfnamefont {J.~M.}\ \bibnamefont {Hamm}}, \
  and\ \bibinfo {author} {\bibfnamefont {O.}~\bibnamefont {Hess}},\ }\href
  {\doibase 10.1103/PhysRevB.97.045428} {\bibfield  {journal} {\bibinfo
  {journal} {Phys. Rev. B}\ }\textbf {\bibinfo {volume} {97}},\ \bibinfo
  {pages} {045428} (\bibinfo {year} {2018})}\BibitemShut {NoStop}%
\bibitem [{\citenamefont {Hamm}\ \emph {et~al.}(2016)\citenamefont {Hamm},
  \citenamefont {Page}, \citenamefont {Bravo-Abad}, \citenamefont
  {Garcia-Vidal},\ and\ \citenamefont {Hess}}]{Hamm}%
  \BibitemOpen
  \bibfield  {author} {\bibinfo {author} {\bibfnamefont {J.~M.}\ \bibnamefont
  {Hamm}}, \bibinfo {author} {\bibfnamefont {A.~F.}\ \bibnamefont {Page}},
  \bibinfo {author} {\bibfnamefont {J.}~\bibnamefont {Bravo-Abad}}, \bibinfo
  {author} {\bibfnamefont {F.~J.}\ \bibnamefont {Garcia-Vidal}}, \ and\
  \bibinfo {author} {\bibfnamefont {O.}~\bibnamefont {Hess}},\ }\href {\doibase
  10.1103/PhysRevB.93.041408} {\bibfield  {journal} {\bibinfo  {journal} {Phys.
  Rev. B}\ }\textbf {\bibinfo {volume} {93}},\ \bibinfo {pages} {041408}
  (\bibinfo {year} {2016})}\BibitemShut {NoStop}%
\bibitem [{\citenamefont {{Das Sarma}}\ and\ \citenamefont
  {Quinn}(1982)}]{DasSarma1982}%
  \BibitemOpen
  \bibfield  {author} {\bibinfo {author} {\bibfnamefont {S.}~\bibnamefont {{Das
  Sarma}}}\ and\ \bibinfo {author} {\bibfnamefont {J.~J.}\ \bibnamefont
  {Quinn}},\ }\href {\doibase 10.1103/PhysRevB.25.7603} {\bibfield  {journal}
  {\bibinfo  {journal} {Phys. Rev. B}\ }\textbf {\bibinfo {volume} {25}},\
  \bibinfo {pages} {7603} (\bibinfo {year} {1982})}\BibitemShut {NoStop}%
\bibitem [{\citenamefont {Vinogradov}\ \emph {et~al.}(2010)\citenamefont
  {Vinogradov}, \citenamefont {Dorofeenko},\ and\ \citenamefont
  {Nechepurenko}}]{A.P.Vinogradov2010}%
  \BibitemOpen
  \bibfield  {author} {\bibinfo {author} {\bibfnamefont {A.}~\bibnamefont
  {Vinogradov}}, \bibinfo {author} {\bibfnamefont {A.}~\bibnamefont
  {Dorofeenko}}, \ and\ \bibinfo {author} {\bibfnamefont {I.}~\bibnamefont
  {Nechepurenko}},\ }\href {\doibase 10.1016/j.metmat.2010.09.002} {\bibfield
  {journal} {\bibinfo  {journal} {Metamaterials}\ }\textbf {\bibinfo {volume}
  {4}},\ \bibinfo {pages} {181} (\bibinfo {year} {2010})}\BibitemShut {NoStop}%
\bibitem [{\citenamefont {Yeh}(2005)}]{Yeh2005}%
  \BibitemOpen
  \bibfield  {author} {\bibinfo {author} {\bibfnamefont {P.}~\bibnamefont
  {Yeh}},\ }\href
  {www.wiley.com/en-us/Optical+Waves+in+Layered+Media-p-9780471731924} {\emph
  {\bibinfo {title} {{Optical waves in layered media}}}}\ (\bibinfo
  {publisher} {Wiley},\ \bibinfo {year} {2005})\BibitemShut {NoStop}%
\bibitem [{\citenamefont {Yeh}\ \emph {et~al.}(1977)\citenamefont {Yeh},
  \citenamefont {Yariv},\ and\ \citenamefont {Hong}}]{Yeh1976}%
  \BibitemOpen
  \bibfield  {author} {\bibinfo {author} {\bibfnamefont {P.}~\bibnamefont
  {Yeh}}, \bibinfo {author} {\bibfnamefont {A.}~\bibnamefont {Yariv}}, \ and\
  \bibinfo {author} {\bibfnamefont {C.-S.}\ \bibnamefont {Hong}},\ }\href
  {\doibase 10.1364/JOSA.67.000423} {\bibfield  {journal} {\bibinfo  {journal}
  {J. Opt. Soc. Am.}\ }\textbf {\bibinfo {volume} {67}},\ \bibinfo {pages}
  {423} (\bibinfo {year} {1977})}\BibitemShut {NoStop}%
\bibitem [{\citenamefont {Gu}\ and\ \citenamefont {Yeh}(1996)}]{Gu1996}%
  \BibitemOpen
  \bibfield  {author} {\bibinfo {author} {\bibfnamefont {C.}~\bibnamefont
  {Gu}}\ and\ \bibinfo {author} {\bibfnamefont {P.}~\bibnamefont {Yeh}},\
  }\href {\doibase 10.1364/OL.21.000504} {\bibfield  {journal} {\bibinfo
  {journal} {Opt. Lett.}\ }\textbf {\bibinfo {volume} {21}},\ \bibinfo {pages}
  {504} (\bibinfo {year} {1996})}\BibitemShut {NoStop}%
\bibitem [{\citenamefont {Agranovich}\ and\ \citenamefont
  {Kravtsov}(1985)}]{Agranovich1985}%
  \BibitemOpen
  \bibfield  {author} {\bibinfo {author} {\bibfnamefont {V.}~\bibnamefont
  {Agranovich}}\ and\ \bibinfo {author} {\bibfnamefont {V.}~\bibnamefont
  {Kravtsov}},\ }\href {\doibase 10.1016/0038-1098(85)91111-1} {\bibfield
  {journal} {\bibinfo  {journal} {Solid State Commun.}\ }\textbf {\bibinfo
  {volume} {55}},\ \bibinfo {pages} {85} (\bibinfo {year} {1985})}\BibitemShut
  {NoStop}%
\bibitem [{\citenamefont {Kidwai}\ \emph {et~al.}(2012)\citenamefont {Kidwai},
  \citenamefont {Zhukovsky},\ and\ \citenamefont {Sipe}}]{Kidwai2012}%
  \BibitemOpen
  \bibfield  {author} {\bibinfo {author} {\bibfnamefont {O.}~\bibnamefont
  {Kidwai}}, \bibinfo {author} {\bibfnamefont {S.~V.}\ \bibnamefont
  {Zhukovsky}}, \ and\ \bibinfo {author} {\bibfnamefont {J.~E.}\ \bibnamefont
  {Sipe}},\ }\href {\doibase 10.1103/PhysRevA.85.053842} {\bibfield  {journal}
  {\bibinfo  {journal} {Phys. Rev. A}\ }\textbf {\bibinfo {volume} {85}},\
  \bibinfo {pages} {053842} (\bibinfo {year} {2012})}\BibitemShut {NoStop}%
\bibitem [{\citenamefont {Soukoulis}\ and\ \citenamefont
  {Wegener}(2010)}]{Soukoulis2010}%
  \BibitemOpen
  \bibfield  {author} {\bibinfo {author} {\bibfnamefont {C.~M.}\ \bibnamefont
  {Soukoulis}}\ and\ \bibinfo {author} {\bibfnamefont {M.}~\bibnamefont
  {Wegener}},\ }\href {\doibase 10.1126/science.1198858} {\bibfield  {journal}
  {\bibinfo  {journal} {Science}\ }\textbf {\bibinfo {volume} {330}},\ \bibinfo
  {pages} {1633} (\bibinfo {year} {2010})}\BibitemShut {NoStop}%
\bibitem [{\citenamefont {Xiao}\ \emph {et~al.}(2010)\citenamefont {Xiao},
  \citenamefont {Drachev}, \citenamefont {Kildishev}, \citenamefont {Ni},
  \citenamefont {Chettiar}, \citenamefont {Yuan},\ and\ \citenamefont
  {Shalaev}}]{Xiao2010}%
  \BibitemOpen
  \bibfield  {author} {\bibinfo {author} {\bibfnamefont {S.}~\bibnamefont
  {Xiao}}, \bibinfo {author} {\bibfnamefont {V.~P.}\ \bibnamefont {Drachev}},
  \bibinfo {author} {\bibfnamefont {A.~V.}\ \bibnamefont {Kildishev}}, \bibinfo
  {author} {\bibfnamefont {X.}~\bibnamefont {Ni}}, \bibinfo {author}
  {\bibfnamefont {U.~K.}\ \bibnamefont {Chettiar}}, \bibinfo {author}
  {\bibfnamefont {H.-K.}\ \bibnamefont {Yuan}}, \ and\ \bibinfo {author}
  {\bibfnamefont {V.~M.}\ \bibnamefont {Shalaev}},\ }\href {\doibase
  10.1038/nature09278} {\bibfield  {journal} {\bibinfo  {journal} {Nature}\
  }\textbf {\bibinfo {volume} {466}},\ \bibinfo {pages} {735} (\bibinfo {year}
  {2010})}\BibitemShut {NoStop}%
\end{thebibliography}
\end{document}